\documentclass[twocolumn, 10pt]{article}

\usepackage[utf8]{inputenc}
\usepackage{amsmath,  amssymb,  amsfonts}
\usepackage{graphicx}
\usepackage{enumitem}
\usepackage{booktabs}
\usepackage[margin=2cm]{geometry}
\usepackage{hyperref}
\usepackage{bm}
\usepackage{cite}
\usepackage{pgfplots}
\pgfplotsset{compat=1.18}
\usepackage{tikz}
\usetikzlibrary{arrows.meta, positioning, fit}
\usepackage{authblk}
\usetikzlibrary{calc}

\title{On the feasibility of model-based feedback control of vertical instability growth rate using out-vessel coils in ARC-like scenarios}
\author[1]{Arunav Kumar}
\author[1]{Cesar Clauser}
\author[1]{Theodore Golfinopoulos}
\affil[1]{Plasma Science and Fusion Center,  Massachusetts Institute of Technology,  Cambridge,  MA 02139,  USA}
\author[2]{Jon C. Hillesheim}
\affil[2]{Commonwealth Fusion Systems, Devens, MA, USA}
\date{(Dated: \today)}

\begin{document}

\twocolumn[
\maketitle

\renewcommand{\thefootnote}{}\footnotetext{\texttt{arunavk@mit.edu}}\renewcommand{\thefootnote}{\arabic{footnote}}

\begin{abstract}
\textcolor{black}{In this work, we propose a model-based feedback controller that regulates the vertical instability growth rate ($\gamma_{gr}$) of a high-elongation, double-null tokamak directly, using only out-vessel poloidal field (PF) coils. High elongation raises the achievable plasma current and fusion performance but makes the plasma vertically unstable, and in a fusion power plant the in-vessel coils that present devices rely on for stabilization may be absent, leaving only distant out-vessel circuits. The controller couples a machine learning surrogate of non-rigid, profile agnostic vertical instability metric to a constrained quadratic program: the surrogate supplies real-time $\gamma_{gr}$ estimates and, via automatic differentiation, the actuator sensitivities, while the program allocates coil voltages to track a target growth rate, maintain double-null divertor balance, and respect electromechanical limits. We tested this method on the ARC~V3A power plant design configuration across 24 closed-loop simulations spanning equilibrium variations, actuator degradations, and transient disturbances. We achieved full or marginal success in 83\% of these cases (full in 50\%, marginal in a further 33\%) and lose control in the remaining 17\%; the failures map the boundary of out-vessel controllability (occurring at the highest growth rates) and under actuator limits. The controller does not regulate boundary shape explicitly: separatrix geometry follows indirectly from growth rate and flux balance control and would require a separate shape control layer for sustained scenario evolution.}
\end{abstract}
\vspace{1em}
]

\section{Introduction}

It is known that high elongation increases the achievable plasma current and fusion performance of a tokamak, but it renders the plasma vertically unstable to the axisymmetric $n=0$ mode, which must be actively stabilized for the discharge to be sustained~\cite{lazarus1990, ariola2016}. The instability is a resistive-wall mode: the characteristic growth time $\tau_\gamma\equiv 1/|\gamma_{gr}|$ is set by the inductive coupling between the plasma and the surrounding conductors, and its value depends not only on elongation but also on the pressure profile and internal inductance~\cite{ward1993}. Loss of vertical control terminates the discharge in a vertical displacement event, driving large conducted heat loads and electromagnetic forces onto the first wall and in-vessel structures~\cite{schuller1995}; avoiding this is a primary constraint on the operating space of any elongated device. Most present tokamaks stabilize the mode with in-vessel coils that provide large bandwidth and passive damping through short coupling distances~\cite{humphreys2007, walker2011, humphreys2009}, and the SPARC tokamak likewise incorporates dedicated in-vessel vertical stability coils~\cite{nelson2024_sparc_vs}. In a fusion power plant, however, the stabilizing coils may have to sit outside the blanket and vacuum vessel: the increased stand-off distance weakens the magnetic coupling and raises the required control effort, and the choice of coil set itself becomes a design variable~\cite{lister1996}. Early feasibility studies showed that out-vessel coils become only marginally controllable at high elongation, particularly under off-normal conditions Nishio \textit{et~al.}\ found that a plasma with $\kappa=2.0$ at low $\beta_p$ and peaked current profile could not be controlled by out-vessel coils within practical voltage limits~\cite{nishio1993}. In double-null (DN) operation the problem is further multivariable: the same poloidal field (PF) actuators that affect vertical stability also modify X-point balance and boundary flux, so vertical stabilization cannot be treated as a decoupled position control loop.

Conventional vertical control regulates the current-centroid position with proportional-derivative (PD) feedback~\cite{humphreys2007, walker2011, ariola2016}. This is effective on many devices, but centroid regulation is not equivalent to regulating the stability margin: the relevant figure of merit is the growth rate itself, which is a function of the field decay index and the plasma internal profiles rather than of vertical position~\cite{portone2005}. A plasma can therefore satisfy a position reference while operating with a dangerously large growth rate, because changes in elongation, internal inductance, or plasma-wall gap that drive $\gamma_{gr}$ are not visible in the centroid displacement. The advantage of regulating $\gamma_{gr}$ directly was recognized on EAST, where model-based growth rate estimates were made available in real time~\cite{wu2021_east, bao2021_east_gamma, detommasi2018}; those implementations used a rigid plasma approximation and served as advisory signals rather than as closed-loop control variables.

Model-based and optimization based controllers  offer a route to closing this loop while respecting actuator limits \cite{Maljaars2017TCV_MPC,Wehner2016DIIID_QProfileMPC,Barton2015DIIID_ActuatorTrajectoryOptimization,Maljaars2015NF_MPCConstraints}. Model predictive control (MPC) has been applied to tokamak magnetic control  of the plasma current and shape, using singular value reduction and online quadratic program (QP) solvers that enforce coil current and voltage constraints explicitly~\cite{gerksic2018}, and to the safety factor profile under time varying constraints~\cite{maljaars2015}. Of particular relevance, explicit MPC has been demonstrated in simulation for vertical stabilization of the $n=0$ mode on ITER~\cite{gerksic2013} and recently in TCV \cite{mele2025experimentaldemonstrationplasmashape}. These approaches share a common structure: a linearized prediction model, a quadratic cost over the controlled outputs, and hard constraints on the actuators that maps naturally onto out-vessel vertical control, where the binding limitation is precisely the voltage authority of the distant coils. What they have lacked is a fast, profile agnostic estimate of the quantity to be regulated.

Machine learning surrogates now supply that estimate. Neural networks trained on equilibrium and stability data can reproduce MHD stability and disruption-related metrics at speeds compatible with real-time control~\cite{kim2024, katesharbeck2019, rea2018}, and physics-informed networks incorporating equilibrium constraints have achieved $<5\%$ error on vertical growth rates across multi-machine datasets~\cite{kim2024}. In parallel, neural-network equilibrium reconstruction now delivers the plasma state at the sub-millisecond rates that feedback requires~\cite{lu2023}, and reinforcement learning controllers have demonstrated combined shape, position, and current control in both simulation and experiment~\cite{degrave2022, seo2026}. Together these advances make direct feedback on $\gamma_{gr}$, using a surrogate fast enough for millisecond control cycles, within reach.

This paper presents a feedback controller that regulates $\gamma_{gr}$ as the primary controlled variable in DN operation with out-vessel actuation, combining the speed of a learned stability surrogate with the constraint-handling of optimization-based control. The approach has three components: (i)~a machine learning surrogate trained on non-rigid stability calculations from the MEQ-FGE-L code suite~\cite{carpanese2020, carpanese_thesis}, which accepts the source profiles $P'$ and $TT'$ (defined in Section~\ref{sec:physics}) as inputs and thereby resolves the profile-aware growth rate variations that integral parameters cannot distinguish; (ii)~automatic differentiation of the surrogate~\cite{hornik1990} to obtain state dependent actuator sensitivities $\partial\gamma_{gr}/\partial\bm{I}_{PF}^{c}$; and (iii)~a constrained quadratic program (QP) that computes coil-voltage commands while coordinating growth rate regulation with X-point flux balance, plasma current, and vertical position under power-supply limits.

\begin{figure}[t]
\centering
\includegraphics[width=\columnwidth]{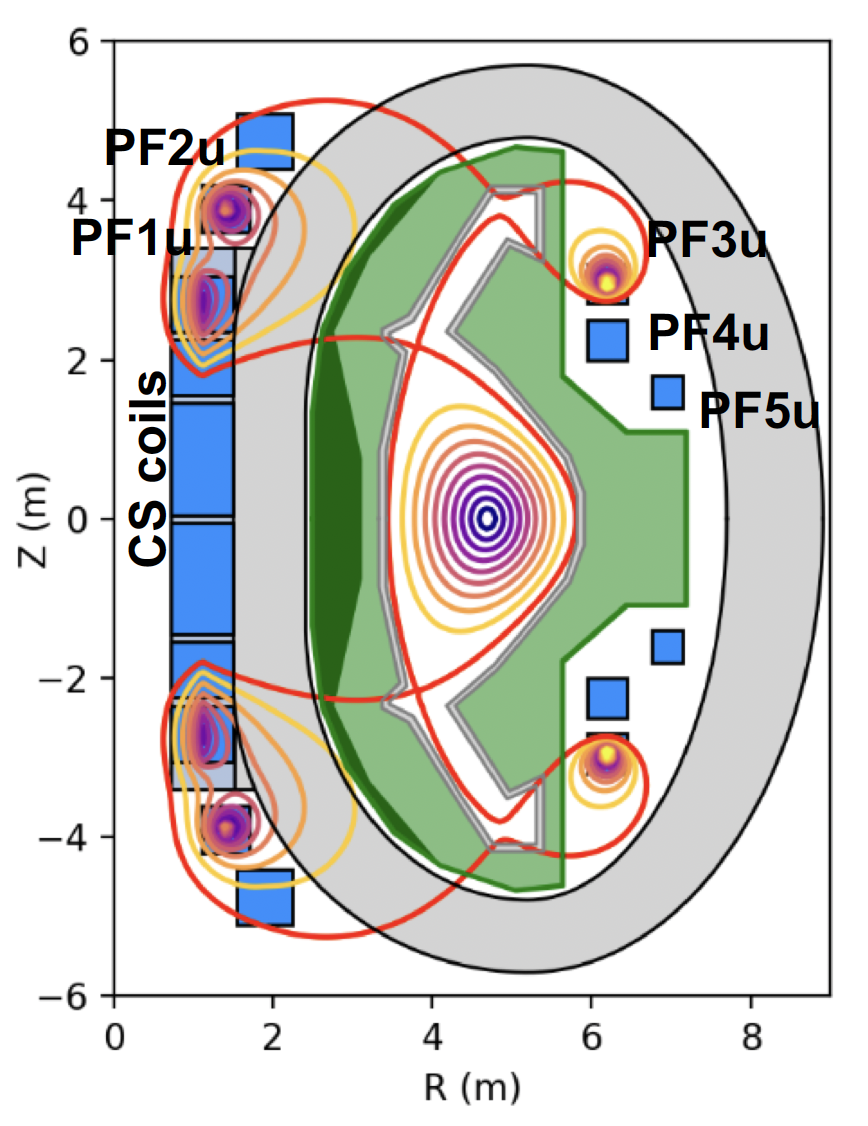}
\caption{Poloidal cross-section and external coil layout for the ARC~V3A configuration used in this work. The TF coil case outline is in grey. The light green is FLiBe, and the dark green is
dedicated neutron shielding.
}
\label{fig:arc_V3A_layout}
\end{figure}

We demonstrate our approach on the ARC~V3A design configuration, a compact high-field fusion power plant that will operate under tighter diagnostic and control constraints than present research devices and is representative of the out-vessel coil control regime~\cite{sorbom2015, rodriguez2024, Creely_2026, Leuthold_2026, Howard_2026, Hillesheim_2026, Sweeney_2026, Eich_2026}. It's H-mode~\cite{wagner1982_hmode} flat-top scenario parameters given as: major radius $R_0=4.62$~m, minor radius $a=1.18$~m, on-axis field $B_0=11.4$~T, plasma current $I_p\simeq 12$~MA, stored energy $W_p=231$~MJ, poloidal beta $\beta_p=0.524$, normalised pressure $\beta_N=1.8$, separatrix triangularity $\delta_{\mathrm{sep}}=0.65$, and target separatrix elongation $\kappa_{\mathrm{sep}}=1.80$ for long flat-top operation ($t_{\mathrm{flattop}}\sim 15$~min); see Hillesheim \textit{et~al.}~\cite{Hillesheim_2026} for a complete overview. For this study, the design assumes no in-blanket or in-vessel magnetic coils; all plasma control must be achieved with coils outside the blanket tank (Fig.~\ref{fig:arc_V3A_layout}) which motivates this work. As the ARC design is still evolving, the analyses here are based on the V3A configuration and are intended to inform future design iterations rather than to characterize a finalized machine.


The scope of this work is limited to vertical stability control. The QP cost function does not include explicit separatrix shape terms (elongation, triangularity, plasma-wall gaps); shape is assumed to be regulated by either a separate outer-loop controller or by the feedforward scenario design, consistent with the timescale separation employed in present tokamak control architectures~\cite{pesamosca2022_tcv, humphreys2007}. The interaction between $\gamma_{gr}$ regulation and boundary shape evolution is discussed in Section~\ref{sec:discussion}. The results are obtained through offline simulation coupling the MEQ equilibrium evolution code with models of coil circuits and power supplies; no experimental validation has been performed, and hardware-in-the-loop and present-day tokamak validation are planned as follow-on work. The purpose here is to establish feasibility, identify where the controller succeeds and where it fails, and provide the basis for that experimental program.

This paper is organized as follows. Section~\ref{sec:physics} formulates the FGEL linearized stability model and describes the ML surrogate architecture and training. Section~\ref{sec:controller} presents the feedback architecture, sensitivity modeling, and QP controller design. Section~\ref{sec:results} reports closed-loop simulation results across 24 test cases for ARC~V3A flat-top scenarios. Section~\ref{sec:discussion} discusses performance and controller boundaries. Section~\ref{sec:conclusions} concludes.

\section{Non-Rigid Growth Rate Estimation and Circuit Dynamics} \label{sec:physics}

\subsection{Computing \texorpdfstring{$\gamma_{gr}$}{gamma\_z} with the FGEL linearized  plasma conductor model}\label{Sec2.1}

Let an equilibrium at time $t$ define a quasi-static operating point satisfying force balance. We write the FGEL linearization of the coupled plasma conductor system about this operating point as
\begin{eqnarray}
    \delta\dot{\mathbf{x}}(t) = \mathbf{A}\delta\mathbf{x}(t) + \mathbf{B}\delta\mathbf{u}(t), 
\end{eqnarray}
where $\mathbf{x}$ collects dynamic states (conductor currents and reduced plasma states),  $\mathbf{u}$ denotes actuator inputs (PF coil voltages),  and $\mathbf{A}$ and $\mathbf{B}$ are the Jacobians of the linearized free-boundary equilibrium evolution coupled to electromagnetic circuit dynamics. \textcolor{black}{This linearized free-boundary model follows the FGE/FGEL formulation of Carpanese~\textit{et al.}~\cite{carpanese2020, carpanese_thesis}. Here and in Eq.~(1) we write $\mathbf{x}$ generically; the reduced dynamic state on which $\mathbf{A}$ and $\mathbf{B}$ act is the $\mathbf{x}_D$ obtained from the static/dynamic partition derived below.}

FGE advances equilibrium evolution by solving a coupled residual at each time step. Two equivalent state representations are commonly used:
\begin{eqnarray}
    \mathbf{x}_\psi = (\psi, \ \mathbf{a}_g, \ \mathbf{I}_e), 
    \qquad
    \mathbf{x}_{I_y} = (\mathbf{I}_y, \ \mathbf{a}_g, \ \mathbf{I}_e), \label{eqn:2}
\end{eqnarray}
where $\psi(R, Z)$ is the poloidal flux,  $\mathbf{I}_y$ is the vector of discretized toroidal plasma current filaments on the grid,  $\mathbf{I}_e$ is the vector of external conductor currents (active PF coils and passive structures),  and $\mathbf{a}_g$ are coefficients of the profile basis functions.

For the ARC~V3A configuration with a $65\times 65$ computational mesh,  the flux based state has dimension $\mathrm{dim}(\mathbf{x}_\psi) = n_\psi + n_g + n_e$,  where $n_\psi = 4225$ (grid points),  $n_g$ is the number of profile basis functions (typically 2 to 4),  and $n_e = n_{\mathrm{PF}} + n_{\mathrm{pass}}$ collects $n_{\mathrm{PF}}=10$ active PF circuits and $n_{\mathrm{pass}}$ passive conductor elements. The dynamic state $\mathbf{x}_D$ has dimension $n_D = n_e + n_g$,  and the system matrix $\mathbf{A}\in\mathbb{R}^{n_D\times n_D}$.

\textcolor{black}{The reduction from the full free-boundary residual to the  state-space form summarized in this subsection follows~\cite{carpanese2020, carpanese_thesis}; we reproduce the key steps for convenience and refer the reader there for full details. The objective is to arrive at the reduced linear system of Eqs.~(14) to (17), whose dominant vertical eigenvalue defines the growth rate $\gamma_{gr}$ that the surrogate is trained to reproduce and the controller regulates.}

\textcolor{black}{Following the MEQ free-boundary formulation~\cite{carpanese2020, carpanese_thesis}, the coupled plasma-conductor system is enforced through the residual}
\begin{eqnarray}
   \mathbf{F}(\mathbf{x}, \dot{\mathbf{x}}, \mathbf{u})=
    \begin{bmatrix}
    \mathbf{F}_p(\mathbf{x})\\
    \mathbf{M}_{ey}\dot{\mathbf{I}}_y+\mathbf{M}_{ee}\dot{\mathbf{I}}_e+\mathbf{R}_e\mathbf{I}_e-\mathbf{V}_e\\
    \boldsymbol{\Lambda}(\mathbf{a}_g, \ldots)
    \end{bmatrix}
    =\mathbf{0}. \label{eq.3}
\end{eqnarray}
Here $\mathbf{F}_p$ is the discretized Grad-Shafranov equilibrium residual. The second block is the external circuit equation,  where $\mathbf{M}_{ee}$ and $\mathbf{M}_{ey}$ are mutual inductance matrices,  $\mathbf{R}_e$ is the conductor resistance matrix,  and $\mathbf{V}_e$ denotes applied voltages to the active coils (passive conductors receive zero applied voltage). The third block $\boldsymbol{\Lambda}$ constrains the profile coefficients (e.g.,  via targets such as $\beta_p$,  $\ell_i$,  $I_p$, $q_{a}$.//),  and can be augmented by current diffusion equations when enabled.

FGE advances this coupled system with an implicit time stepping scheme. FGEL is obtained by linearizing the evolution operator around a converged equilibrium:
\begin{eqnarray}
    \mathbf{F}(\mathbf{x}, \dot{\mathbf{x}}, \mathbf{u})=\mathbf{0}, 
\end{eqnarray}
where $\mathbf{u}$ collects control inputs. The residual is split into static and dynamic parts, 
\begin{eqnarray}
\mathbf{F}=\mathbf{F}_S(\mathbf{x}, \mathbf{u})+\mathbf{F}_D(\mathbf{x}, \dot{\mathbf{x}}, \mathbf{u}), 
\end{eqnarray}
based on whether terms depend on $\dot{\mathbf{x}}$. The state is split as
\begin{eqnarray}
    \mathbf{x}=
    \begin{bmatrix}
    \mathbf{x}_S\\
    \mathbf{x}_D
    \end{bmatrix}, 
\end{eqnarray}
\textcolor{black}{where the full parameter state of Eq.~(2) is partitioned so that $\mathbf{x}_S$ groups the algebraic (force balance) variables and $\mathbf{x}_D$ groups the dynamic variables (conductor current states and any dynamic profile states). The split is chosen so that $\partial \mathbf{F}_S/\partial \mathbf{x}_S$ is full rank; eliminating $\mathbf{x}_S$ then leaves $\mathbf{x}_D$ as the reduced state of Eq.~(1).}

Let $(\mathbf{x}_0, \mathbf{u}_0)$ be a converged equilibrium with $\dot{\mathbf{x}}_0=\mathbf{0}$. For small perturbations $\delta\mathbf{x}$ and $\delta\mathbf{u}$,  the first-order expansion yields
\begin{eqnarray}
    \mathbf{0}=
    \frac{\partial \mathbf{F}_S}{\partial \mathbf{x}_S}\delta\mathbf{x}_S+
    \frac{\partial \mathbf{F}_S}{\partial \mathbf{x}_D}\delta\mathbf{x}_D+
    \frac{\partial \mathbf{F}_S}{\partial \mathbf{u}}\delta\mathbf{u}, 
\end{eqnarray}
\begin{multline}
    \mathbf{0}=
    \frac{\partial \mathbf{F}_D}{\partial \dot{\mathbf{x}}_S}\delta\dot{\mathbf{x}}_S+
    \frac{\partial \mathbf{F}_D}{\partial \dot{\mathbf{x}}_D}\delta\dot{\mathbf{x}}_D \\
    +\frac{\partial \mathbf{F}_D}{\partial \mathbf{x}_S}\delta\mathbf{x}_S+
    \frac{\partial \mathbf{F}_D}{\partial \mathbf{x}_D}\delta\mathbf{x}_D+
    \frac{\partial \mathbf{F}_D}{\partial \mathbf{u}}\delta\mathbf{u}.
\end{multline}
All Jacobians are evaluated at $(\mathbf{x}_0, \mathbf{u}_0)$. The first equation eliminates $\delta\mathbf{x}_S$:
\begin{eqnarray}
    \delta\mathbf{x}_S=
    \frac{\partial \mathbf{x}_S}{\partial \mathbf{x}_D}\delta\mathbf{x}_D+
    \frac{\partial \mathbf{x}_S}{\partial \mathbf{u}}\delta\mathbf{u}, 
\end{eqnarray}
\begin{eqnarray}
    \frac{\partial \mathbf{x}_S}{\partial \mathbf{x}_D}:=
    -\left(\frac{\partial \mathbf{F}_S}{\partial \mathbf{x}_S}\right)^{-1}\frac{\partial \mathbf{F}_S}{\partial \mathbf{x}_D}, \\
    \quad
    \frac{\partial \mathbf{x}_S}{\partial \mathbf{u}}:=
    -\left(\frac{\partial \mathbf{F}_S}{\partial \mathbf{x}_S}\right)^{-1}\frac{\partial \mathbf{F}_S}{\partial \mathbf{u}}.
\end{eqnarray}
Substitution gives a reduced parameter form for the dynamic perturbations:
\begin{eqnarray}
    \mathbf{0}=\mathbf{S}\delta\dot{\mathbf{x}}_D-\mathbf{K}\delta\mathbf{x}_D-\mathbf{U}\delta\mathbf{u}-\mathbf{V}\delta\dot{\mathbf{u}}, 
\end{eqnarray}
with
\begin{align}
    \mathbf{S} &:=\frac{\partial \mathbf{F}_D}{\partial \dot{\mathbf{x}}_D}+\frac{\partial \mathbf{F}_D}{\partial \dot{\mathbf{x}}_S}\frac{\partial \mathbf{x}_S}{\partial \mathbf{x}_D}, \quad
    \mathbf{K} :=-\frac{\partial \mathbf{F}_D}{\partial \mathbf{x}_D}-\frac{\partial \mathbf{F}_D}{\partial \mathbf{x}_S}\frac{\partial \mathbf{x}_S}{\partial \mathbf{x}_D},  \nonumber \\
    \mathbf{U} &:=-\frac{\partial \mathbf{F}_D}{\partial \mathbf{u}}-\frac{\partial \mathbf{F}_D}{\partial \mathbf{x}_S}\frac{\partial \mathbf{x}_S}{\partial \mathbf{u}}, \quad
    \mathbf{V} :=-\frac{\partial \mathbf{F}_D}{\partial \dot{\mathbf{x}}_S}\frac{\partial \mathbf{x}_S}{\partial \mathbf{u}}.
\end{align}
If $\mathbf{S}$ is invertible,  FGEL can be written as a standard state-space model:
\begin{eqnarray}
    \delta\dot{\mathbf{x}}_D=\mathbf{A}\delta\mathbf{x}_D+\mathbf{B}\delta\hat{\mathbf{u}}, 
    \qquad
    \delta\hat{\mathbf{u}}:=\begin{bmatrix}\delta\mathbf{u}\\ \delta\dot{\mathbf{u}}\end{bmatrix}, 
\end{eqnarray}
\begin{eqnarray}
\mathbf{A}=\mathbf{S}^{-1}\mathbf{K}, 
    \qquad
\mathbf{B}=\big[\mathbf{S}^{-1}\mathbf{U}\ \ \mathbf{S}^{-1}\mathbf{V}\big].
\end{eqnarray}
In the unforced case ($\delta\hat{\mathbf{u}}=\mathbf{0}$),  the perturbations evolve as
\begin{eqnarray}
\delta\mathbf{x}_D(t)=\sum_i \alpha_i \mathbf{v}_i e^{\lambda_i t}, 
    \qquad
\mathbf{A}\mathbf{v}_i=\lambda_i\mathbf{v}_i, 
\end{eqnarray}
so $\Re(\lambda_i)$ is the exponential growth/decay rate of mode $i$ (in  $\mathrm{s}^{-1}$). We define the vertical growth rate as
\begin{eqnarray}
    \gamma_{gr} \equiv \Re(\lambda_z), 
\end{eqnarray}
where $\lambda_z$ is the eigenvalue associated with the dominant vertical eigenmode. In practice,  $\lambda_z$ is identified using a vertical observable to select the eigenmode with the strongest vertical signature. This $\gamma_{gr}$ is the quantity regulated by the proposed controller. Related work has focused on computationally efficient vertical instability calculations suitable for real-time applications~\cite{olofsson2022_fast_vertical}.

We note that rigid approximation treats the plasma current distribution as fixed under small perturbations. While adequate at modest elongation,  in DN high $\kappa$ regimes the vertical branch is sensitive to internal inductance $\ell_i$,  pressure profile shape,  and the plasma-wall gap,  all of which change the effective inductive coupling between the plasma and external conductors. Rigid models can therefore exhibit systematic bias in $\gamma_{gr}$ and in the inferred stability margin. \textcolor{black}{Non-rigid plasma response frameworks retain profile degrees of freedom and compute consistent Jacobians for force balance, which yields more reliable $\gamma_{gr}$ estimates in high $\kappa$ regimes.}

\subsection{Neural network surrogate: data,  architecture,  and training}
\label{SEC.2.2}
We now describe the dataset,  network architecture,  and training objective.

\paragraph{Database.}
The training database was constructed from the ARC V3A baseline
flat-top equilibrium~\cite{Hillesheim_2026}.
This reference was augmented through Latin Hypercube Sampling, 
varying plasma current
$I_p \in [8, 12]\, \mathrm{MA}$, 
areal elongation
$\kappa_{\mathrm{areal}} \in [1.5, 1.95]$, 
triangularity
$\delta \in [0.3, 0.5]$, 
the coil current combinations, 
\textcolor{black}{and the $P'$ and $TT'$ profile shapes (sampled through internal inductance $\ell_i \in [0.5, 0.9]$ and pedestal height/width, so that the profiles are not held fixed across the database)}, 
yielding approximately $6{, }000$ labeled equilibria.
Each sample corresponds to an independent Grad-Shafranov solve
with FGEL linearization (Section~\ref{Sec2.1}) at the perturbed operating point.
The dataset was split into training (70\%),  validation (15\%), 
and test (15\%) subsets,  stratified to ensure uniform coverage
of the sampled parameter ranges.

\textcolor{black}{For each of the $\sim$6000 equilibria in the database,  FGEL linearization (Section~\ref{Sec2.1}) was performed,  and the vertical growth rate $\gamma_{gr}$,  perturbed poloidal flux $\delta\psi(R_k, Z_k)$,  and perturbed toroidal current density $\delta j_\phi(R_k, Z_k)$ on a set of $N_g$ grid points were stored as labels,  prior to the train/validation/test split.}

\paragraph{Input representation.}
The surrogate accepts two categories of input.

\begin{enumerate}
\item \textit{Scalar features} ($\mathbf{s}_{\rm sc}\in\mathbb{R}^{25}$): plasma current $I_p$,  volume-averaged electron density $\langle n_e \rangle$,  stored energy $W_{J}$,  inner and outer plasma-wall gaps $g_{\text{in}}$ and $g_{\text{out}}$,  vertical current centroid $z_{I_p}$,  areal elongation $\kappa_{\text{areal}}$,  internal inductance $\ell_i$,  on-axis toroidal field $B_0$,  poloidal field coil currents $\mathbf{I}_{PF}^{c} \in \mathbb{R}^{10}$,  primary and secondary X-point coordinates $(R_X,  Z_X)_{p, s}$,  and flux at two fixed control points $c_1,  c_2$. All scalars are normalized to zero mean and unit variance using training set statistics. \textcolor{black}{These scalars are the global equilibrium parameters,  actuator currents,  and boundary/X-point geometry that enter the axisymmetric vertical stability problem. A feature relevance check (permutation importance on the held-out set) found $\gamma_{gr}$ accuracy dominated by $I_p$,  $\ell_i$,  $\kappa_{\mathrm{areal}}$,  and the PF currents; the remaining features contribute primarily to the auxiliary $\delta\psi$ and $\delta j_\phi$ outputs and are retained for that reason.}

\item \textit{Profiles.} \textcolor{black}{$P'(\psi)\equiv \mathrm{d}p/\mathrm{d}\psi$ is the pressure gradient and $TT'(\psi)$ is the toroidal field function term,  where $T(\psi)\equiv R B_\phi$ is the toroidal field function (denoted $F$ or $g$ elsewhere in the literature),  so that $TT' = F F'$.} Both are provided on a uniform normalized poloidal flux grid $\psi_N \in [0, 1]$ with $n_\psi = 65$ points. Together they encode the pressure-gradient and toroidal field function contributions to the equilibrium current density:
\begin{eqnarray}
    j_\phi(R, Z) = R\, P'(\psi) + \frac{1}{\mu_0 R}\, TT'(\psi), 
    \label{eq:GS_source}
\end{eqnarray}
and are one of the key drivers of the non-rigid plasma response. In particular,  the gradient structure of $P'$ near the pedestal and $TT'$ at mid-radius strongly influences internal inductance and hence the vertical instability~\cite{carpanese_thesis}. Including profiles as inputs rather than relying solely on integral quantities such as $\ell_i$ and $\beta_N$ allows the surrogate to resolve profile dependent variations in $\gamma_{gr}$ that integral parameters cannot distinguish.
\end{enumerate}

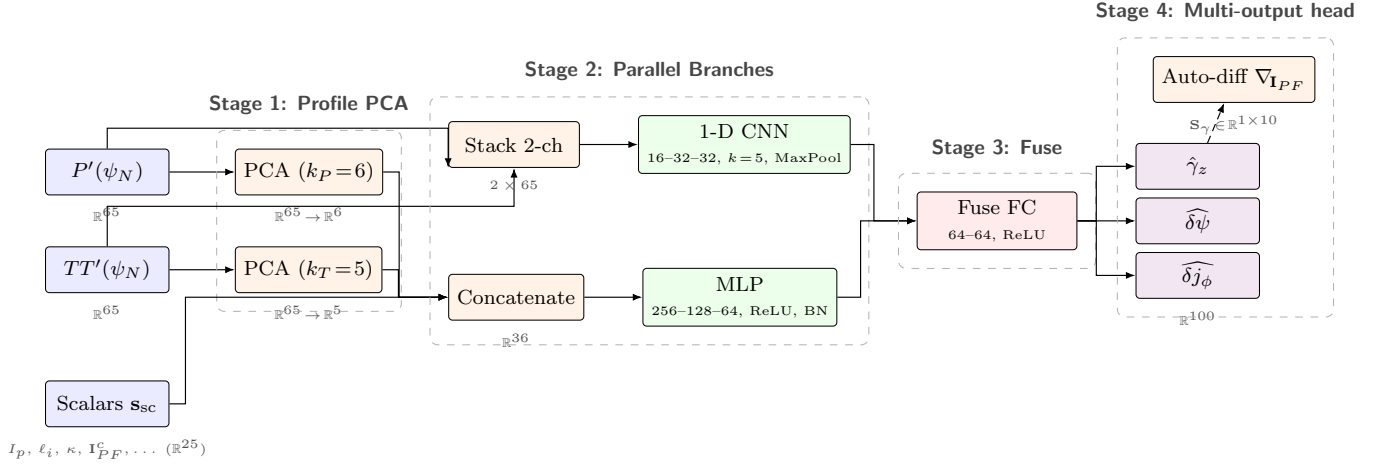
\begin{figure*}[t]
\centering
\resizebox{0.65\paperheight}{!}{


\begin{tikzpicture}[
  auto,
  node distance=3.5mm and 5.5mm,
  input/.style={
    draw, rectangle, rounded corners=1.5pt,
    minimum height=6.3mm, minimum width=17mm,
    align=center, font=\footnotesize,
    fill=blue!8},
  proc/.style={
    draw, rectangle, rounded corners=1.5pt,
    minimum height=6.3mm, minimum width=18mm,
    align=center, font=\footnotesize,
    fill=orange!10},
  branch/.style={
    draw, rectangle, rounded corners=1.5pt,
    minimum height=6.3mm, minimum width=22mm,
    align=center, font=\footnotesize,
    fill=green!8},
  fusion/.style={
    draw, rectangle, rounded corners=1.5pt,
    minimum height=6.3mm, minimum width=22mm,
    align=center, font=\footnotesize,
    fill=red!8},
  output/.style={
    draw, rectangle, rounded corners=1.5pt,
    minimum height=6.3mm, minimum width=17mm,
    align=center, font=\footnotesize,
    fill=violet!10},
  dimlab/.style={font=\tiny\sffamily, text=gray!65!black},
  signal/.style={-{Latex[scale=0.75]}, line width=0.45pt},
  dashed signal/.style={-{Latex[scale=0.75]}, line width=0.45pt, densely dashed},
  stagebox/.style={draw, dashed, rounded corners=3pt, inner sep=2.5mm,line width=0.4pt, gray!60},
  stagelabel/.style={font=\footnotesize\bfseries\sffamily, text=gray!55!black},
]

\node[input] (pprime) {$P'(\psi_N)$};
\node[dimlab, below=0.5mm of pprime] {$\mathbb{R}^{65}$};

\node[input, below=7mm of pprime] (ttprime) {$TT'(\psi_N)$};
\node[dimlab, below=0.5mm of ttprime] {$\mathbb{R}^{65}$};

\node[input, below=12mm of ttprime] (scalars) {Scalars $\mathbf{s}_{\rm sc}$};
\node[dimlab, below=0.5mm of scalars]
  {$I_p,\,\ell_i,\,\kappa,\,\mathbf{I}_{PF}^c,\ldots\;(\mathbb{R}^{25})$};

\node[proc, right=9mm of pprime] (pca_p) {PCA ($k_P\!=\!6$)};
\node[dimlab, below=0.5mm of pca_p] {$\mathbb{R}^{65}\!\to\!\mathbb{R}^{6}$};

\node[proc, right=9mm of ttprime] (pca_t) {PCA ($k_T\!=\!5$)};
\node[dimlab, below=0.5mm of pca_t] {$\mathbb{R}^{65}\!\to\!\mathbb{R}^{5}$};

\draw[signal] (pprime) -- (pca_p);
\draw[signal] (ttprime) -- (pca_t);

\node[stagebox, fit=(pca_p)(pca_t)] (box1) {};
\node[stagelabel, anchor=south] at ($(box1.north)+(0,1.0mm)$) {Stage 1: Profile PCA};

\coordinate (stage2_anchor) at ($(pca_p.east)!0.5!(pca_t.east)$);

\node[proc, right=9mm of stage2_anchor, yshift=10.5mm] (stack) {Stack 2-ch};
\node[dimlab, below=0.5mm of stack] {$2\times 65$};

\draw[signal] (pprime.north) |- ([yshift=2.8mm]pprime.north) -| (stack.south west);
\draw[signal] (ttprime.north) |- ([yshift=5.5mm]ttprime.north east) -| (stack.south);

\node[branch, right=8mm of stack] (cnn)
  {1-D CNN\\{\tiny 16--32--32, $k\!=\!5$, MaxPool}};
\draw[signal] (stack) -- (cnn);

\node[proc, right=9mm of stage2_anchor, yshift=-10.5mm] (concat) {Concatenate};
\node[dimlab, below=0.5mm of concat] {$\mathbb{R}^{36}$};

\draw[signal] (pca_p.east) -| ([xshift=2.2mm]pca_p.east) |- (concat.west);
\draw[signal] (pca_t.east) -| ([xshift=2.2mm]pca_t.east) |- (concat.west);
\draw[signal] (scalars.east) -| ([xshift=2.2mm]scalars.east) |- (concat.west);

\node[branch, right=8mm of concat] (mlp)
  {MLP\\{\tiny 256--128--64, ReLU, BN}};
\draw[signal] (concat) -- (mlp);

\node[stagebox, fit=(stack)(cnn)(concat)(mlp)] (box2) {};
\node[stagelabel, anchor=south] at ($(box2.north)+(0,1.0mm)$) {Stage 2: Parallel Branches};

\coordinate (mid_branches) at ($(cnn.east)!0.5!(mlp.east)$);
\node[fusion, right=10mm of mid_branches] (fuse)
  {Fuse FC\\{\tiny 64--64, ReLU}};

\draw[signal] (cnn.east) -- ++(3.2mm,0) |- (fuse.west);
\draw[signal] (mlp.east) -- ++(3.2mm,0) |- (fuse.west);

\node[stagebox, fit=(fuse)] (box3) {};
\node[stagelabel, anchor=north] at ($(box3.north)+(0,5.8mm)$) {Stage 3: Fuse};

\node[output, right=8mm of fuse, yshift=7.5mm] (outg) {$\hat{\gamma}_z$};
\node[dimlab, below=0.5mm of outg] {$\mathbb{R}^{1}$};

\node[output, right=8mm of fuse] (outpsi) {$\widehat{\delta\psi}$};
\node[dimlab, below=0.5mm of outpsi] {$\mathbb{R}^{100}$};

\node[output, right=8mm of fuse, yshift=-7.5mm] (outj) {$\widehat{\delta j_\phi}$};
\node[dimlab, below=0.5mm of outj] {$\mathbb{R}^{100}$};

\draw[signal] (fuse.east) -- ++(2.6mm,0) |- (outg.west);
\draw[signal] (fuse.east) -- (outpsi.west);
\draw[signal] (fuse.east) -- ++(2.6mm,0) |- (outj.west);

\node[proc, above=5.5mm of outg, xshift=5mm, minimum width=20mm] (autodiff)
  {Auto-diff $\nabla_{\!\mathbf{I}_{PF}}$};
\node[dimlab, below=0.5mm of autodiff] {$\mathbf{S}_\gamma\!\in\!\mathbb{R}^{1\times 10}$};
\draw[dashed signal] (outg) -- (autodiff);
\node[stagebox, fit=(outg)(outpsi)(outj)(autodiff)] (box4) {};
\node[stagelabel, anchor=south] at ($(box4.north)+(0,1.0mm)$) {Stage 4: Multi-output head};
\end{tikzpicture}}
\caption{Schematic of the MEQ-ML-DN surrogate architecture. The profile branch compresses $P'(\psi_N)$ and $TT'(\psi_N)$ via PCA and a 1D convolutional feature extractor; the scalar branch processes equilibrium parameters and coil currents through a fully connected network. Both branches are fused and mapped to the multi-output head.}
\label{fig:ml_model}
\end{figure*}

\paragraph{Architecture.}
The surrogate employs a dual-branch architecture that processes scalar and profile inputs through separate pathways before fusion,  following the multimodal paradigm established for tokamak stability surrogates~\cite{seo2026, kim2024,  holt2024, rothstein2021neural}. The ML architecture,  shown schematically in Fig.~\ref{fig:ml_model},  is described as follows.

\textit{Stage~1: Profile compression via \textcolor{black}{principal component analysis (PCA)}.}
Each source-function profile is projected onto a truncated PCA basis:
\begin{eqnarray}
    \mathbf{p}' \approx \bar{\mathbf{p}}' +
    \sum_{m=1}^{k_P} \alpha_m^{(P)} \mathbf{e}_m^{(P)}, 
    \label{eq:pca_Pprime}
\end{eqnarray}
where $\bar{\mathbf{p}}'$ is the training-set mean,  $\{\mathbf{e}_m^{(P)}\}$ are orthonormal eigenvectors,  and $\alpha_m^{(P)}$ are projection coefficients. Analogously for $TT'$. The number of retained modes satisfies
\begin{eqnarray}
    \frac{\sum_{m=1}^{k_P} \sigma_m^2}{\sum_{m=1}^{n_\psi} \sigma_m^2}
    \geq 0.995, 
    \label{eq:pca_variance}
\end{eqnarray}
with $k_P = 6$ and $k_T = 5$ (99.7\% and 99.5\% variance,  respectively),  \textcolor{black}{the criterion in Eq.~\eqref{eq:pca_variance} being applied separately to each profile. This reduces the combined profile dimensionality from $130 = 2\,n_\psi$ (two profiles $P'$ and $TT'$,  each on $n_\psi=65$ points) to $k_P+k_T=11$ PCA coefficients.}

\textit{Stage~2: Parallel processing branches.}
The scalar \textcolor{black}{multilayer perceptron (MLP)}\cite{rumelhart1986, goodfellow2016} branch concatenates the 25 normalized scalar features with the 11 PCA coefficients. This produces 36 dimensional input processed through three hidden layers (256-128-64 neurons,  ReLU activations ~\cite{nair2010, glorot2011},  batch normalization ~\cite{ioffe2015}) to yield an embedding $\mathbf{h}_{\rm MLP} \in \mathbb{R}^{64}$. The \textcolor{black}{convoulational neural network}~\cite{lecun1998, krizhevsky2012} \textcolor{black}{(CNN)} branch stacks the raw $P'$ and $TT'$ profiles into a two-channel 1D tensor of length~65 and processes it through three convolutional blocks (16,  32,  32 filters; kernel size 5; pooling) to produce $\mathbf{h}_{\rm CNN} \in \mathbb{R}^{64}$. The MLP captures global nonlinear relationships among integrated quantities and PCA compressed profiles,  while the CNN extracts spatially localized features such as pedestal steepness and inflection points directly from the profile waveforms ~\cite{kiranyaz2021}.

\textit{Stage~3: Fuse.}
The branch embeddings are concatenated,  $\mathbf{h}_{\rm cat} = [\mathbf{h}_{\rm MLP};\ \mathbf{h}_{\rm CNN}] \in \mathbb{R}^{128}$,  and processed through two fully connected layers (64 neurons each,  ReLU) to produce $\mathbf{h}_{\rm fused} \in \mathbb{R}^{64}$.

\textit{Stage~4: Output head.}
A final linear layer maps $\mathbf{h}_{\rm fused}$ to three outputs:
\begin{eqnarray}
    \hat{\gamma}_z, \qquad
    \widehat{\delta\psi}(R_k,  Z_k) \in \mathbb{R}^{N_g}, \qquad
    \widehat{\delta j_\phi}(R_k,  Z_k) \in \mathbb{R}^{N_g}, 
    \label{eq:surrogate_outputs}
\end{eqnarray}
where $N_g = 60$ selected grid points span the plasma cross-section. The spatial outputs are included not because they are needed for the $\gamma_{gr}$ feedback loop,  but because they enforce physically consistent internal representations and enable future coupling to shape control modules.

\paragraph{Loss function.}
Training minimizes a composite loss:
\begin{eqnarray}
\begin{split}
    \mathcal{L} = \frac{1}{N}\sum_{i=1}^{N}\Big[
    &\alpha_1\big(\gamma_{gr}^{(i)}-\hat{\gamma}_z^{(i)}\big)^2 \\
    &+ \frac{\alpha_2}{N_g}
      \big\|\delta\psi^{(i)}-\widehat{\delta\psi}^{(i)}\big\|_2^2 \\
    &+ \frac{\alpha_3}{N_g}
      \big\|\delta j_\phi^{(i)}
      - \widehat{\delta j_\phi}^{(i)}\big\|_2^2\Big]
    + \lambda\sum_l \|\mathbf{W}_l\|_F^2, 
\end{split}
\label{eq:loss_total}
\end{eqnarray}
with weights $\alpha_1 = 1.0$,  $\alpha_2 = 0.1$,  $\alpha_3 = 0.1$ which give priority to growth rate accuracy. Each spatial term is normalized by $N_g$ to prevent the higher dimensional field outputs from dominating. The final term applies L2 weight decay ($\lambda = 10^{-4}$). Ablation studies (not shown here) confirm that including flux and current density terms improves $\gamma_{gr}$ generalization (see Kumar \textit{et al.} \cite{Kumar_2026}) by approximately 15\% compared to a single output model.

Training used the Adam optimizer~\cite{kingma2015adam} with initial learning rate $10^{-3}$; it decayed by 0.5 every 25 epochs; batches were of 128 samples,  for 60 epochs ($\approx$2~hours on an NVIDIA Tesla V60 GPU).
We implement early stopping with  patience of 15 epochs.

\begin{table*}[ht]
\centering
\caption{MEQ-ML-DN surrogate performance on held-out test set (900 equilibria). The unoptimized inference time (0.9~ms) is the conservative single-core benchmark; runtime-optimized inference (0.4~ms) is used in the closed-loop timing budget (Section~\ref{sec:discussion}).}
\label{tab:surrogate_performance}
\begin{tabular}{lcccccc}
\hline
Dataset & MAE ($\gamma_{gr}$) [s$^{-1}$]
        & RMSE [s$^{-1}$]
        & $R^2$
        & Inference (unopt.) [ms]
        & Inference (opt.) [ms] \\
\hline
ARC~V3A & 1.8 $\pm$ 0.2
        & 2.4 $\pm$ 0.3
        & 0.972
        & 0.9 $\pm$ 0.1
        & 0.4 $\pm$ 0.05 \\
\hline
\end{tabular}
\end{table*}

Table~\ref{tab:surrogate_performance} summarizes validation metrics on the held-out test set (900 equilibria). The surrogate attains a mean absolute error of 1.8~s$^{-1}$ for $\gamma_{gr}$,  corresponding to a relative error of 4.2\%.
Errors are largest at the boundaries of the sampled parameter space,  particularly at high $\kappa_{\mathrm{areal}}$ combined with low $\ell_i$ or extreme coil current combinations,  where the surrogate extrapolates beyond densely sampled regions. The quasi static assumption underlying MEQ-FGE-L is appropriate for the flat-top conditions studied here; extension to current ramp phases,  where $|\dot{I}_p|$ can exceed 0.5~MA/s,  would require time dependent training data and is left to future work.

Inference time on a single Intel Xeon CPU core (2.4~GHz) averages $0.9 \pm 0.1$~ms without graph optimization. With runtime optimization (TorchScript or ONNX \cite{onnx} export),  inference drops to approximately 0.4~ms,  consistent with the requirements of 1~kHz control cycles.

\section{Feedback Architecture via ML Surrogates and QP}\label{sec:controller}


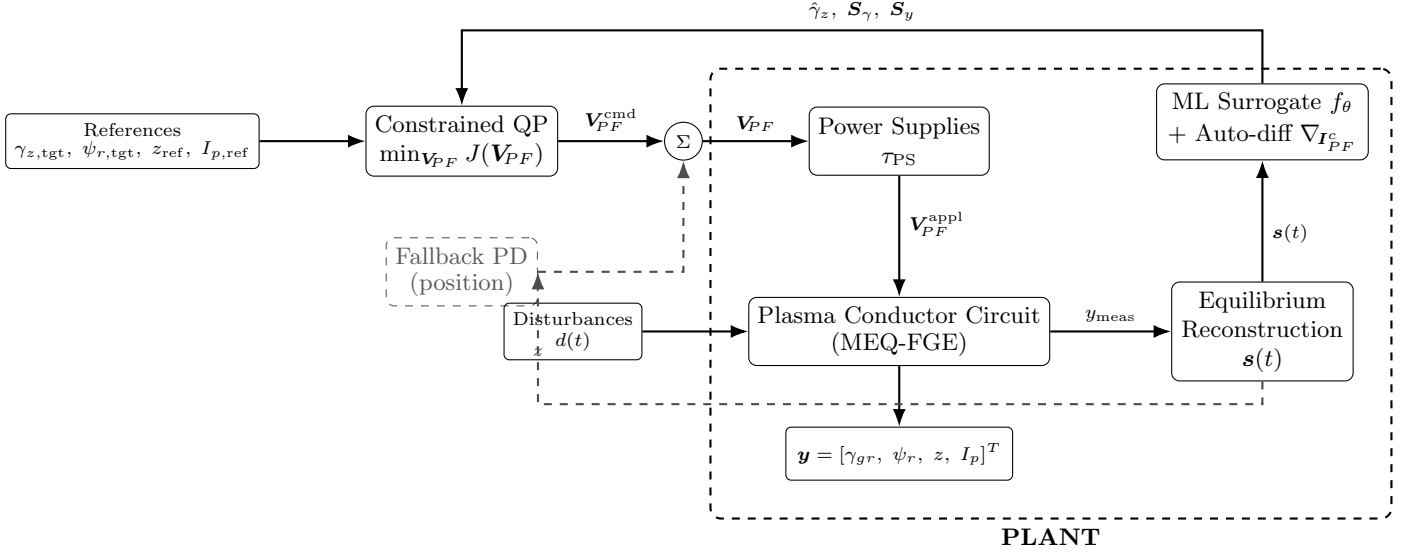
\begin{figure*}[t]
\centering
\begin{tikzpicture}[
  auto, 
  >=Latex, 
  node distance=10mm, 
  block/.style={draw,  rectangle,  rounded corners=3pt,  minimum height=9mm, 
                minimum width=20mm,  align=center,  font=\small}, 
  smallblock/.style={draw,  rectangle,  rounded corners=2pt,  minimum height=7mm, 
                     minimum width=16mm,  align=center,  font=\scriptsize}, 
  dashedblock/.style={draw,  dashed,  rectangle,  rounded corners=3pt, 
                      minimum height=9mm,  minimum width=20mm,  align=center, 
                      font=\small,  gray!70!black}, 
  sum/.style={draw,  circle,  inner sep=0pt,  minimum size=5mm,  font=\scriptsize}, 
  signal/.style={-Latex,  thick}, 
  dsignal/.style={-Latex,  thick,  dashed,  gray!60!black}, 
  dashedbox/.style={draw,  dashed,  rounded corners=4pt,  inner sep=5mm,  thick}, 
  outerbox/.style={draw,  densely dotted,  rounded corners=4pt,  inner sep=4mm, 
                   thick,  gray!50!black}
]

\node[smallblock] (ref) {References\\[-1pt]
  $\gamma_{z, \mathrm{tgt}}, \;\psi_{r, \mathrm{tgt}}, \;z_{\mathrm{ref}}, \;I_{p, \mathrm{ref}}$};

\node[block,  right=14mm of ref] (qp) {Constrained QP\\[0pt]
  $\min_{\bm{V}_{\!PF}} J(\bm{V}_{\!PF})$};

\node[dashedblock,  below=8mm of qp] (fallback)
  {Fallback PD\\[-1pt] (position)};

\node[sum,  right=14mm of qp] (sw) {$\Sigma$};

\node[block,  right=14mm of sw,  minimum width=22mm] (ps)
  {Power Supplies\\$\tau_{\mathrm{PS}}$};

\node[block,  below=16mm of ps,  minimum width=38mm] (plant)
  {Plasma Conductor Circuit\\[-1pt]
    (MEQ-FGE)};

\node[smallblock,  left=14mm of plant] (dist) {Disturbances\\$d(t)$};

\node[block,  right=16mm of plant,  minimum width=24mm] (diag)
  {Equilibrium\\Reconstruction\\$\bm{s}(t)$};

\node[block,  above=16mm of diag,  minimum width=28mm] (ml)
  {ML Surrogate $f_\theta$\\[1pt]
   + Auto-diff $\nabla_{\!\bm{I}_{PF}^{c}}$};

\node[smallblock,  below=8mm of plant,  minimum width=30mm] (yout)
  {$\bm{y}=[\gamma_{gr}, \;\psi_r, \;z, \;I_p]^T$};

\draw[signal] (ref) -- (qp);
\draw[signal] (qp) -- node[above,  font=\scriptsize] {$\bm{V}_{\!PF}^{\rm cmd}$} (sw);
\draw[dsignal] (fallback) -| (sw);
\draw[signal] (sw) -- node[above,  font=\scriptsize] {$\bm{V}_{\!PF}$} (ps);
\draw[signal] (ps) -- node[right,  font=\scriptsize,  pos=0.4] {$\bm{V}_{\!PF}^{\rm appl}$} (plant);
\draw[signal] (dist) -- (plant);
\draw[signal] (plant) -- (yout);
\draw[signal] (plant) -- node[above,  font=\scriptsize] {$y_{\rm meas}$} (diag);
\draw[signal] (diag) -- node[right,  font=\scriptsize,  pos=0.4] {$\bm{s}(t)$} (ml);
\draw[signal] (ml.north) |- ++(0, 7mm) -|
  node[pos=0.25,  above,  font=\scriptsize] {$\hat{\gamma}_z, \;\bm{S}_\gamma, \;\bm{S}_y$}
  (qp.north);
\draw[dsignal] (diag.south) |- ++(0, -3mm) -|
  node[pos=0.75,  below,  font=\scriptsize,  gray!60!black] {$z$}
  (fallback.east);

\node[dashedbox,  fit=(ps)(plant)(diag)(yout), 
      label={[font=\bfseries\small,  anchor=north]below:PLANT}] (pbox) {};

\end{tikzpicture}
\caption{Control architecture for vertical instability growth rate regulation.
The inner loop (this work) uses the ML surrogate to estimate
$\hat{\gamma}_z$ and actuator sensitivities $\bm{S}_\gamma$,  which feed a
constrained QP that commands PF coil voltages. Controlled outputs are
$\bm{y}=[\gamma_{gr}, \, \psi_r, \, z, \, I_p]^T$; boundary shape parameters
$(\kappa, \delta, \mathrm{gaps})$ are not directly included in the QP cost. A dashed
fallback PD position controller indicates an architectural element
required for integrated deployment but not implemented here.}
\label{fig:plant_model_block}
\end{figure*}

\textcolor{black}{The overall control architecture is shown in Fig.~\ref{fig:plant_model_block}: the ML surrogate and its automatic differentiation sensitivities feed a constrained QP that commands the PF voltages,  with the plant closed through the MEQ-FGE circuit model and equilibrium reconstruction.} 

\textcolor{black}{Let $\mathbf{s}(t)$ denote the equilibrium state vector available at runtime:
\begin{eqnarray}
    \mathbf{s} \equiv [I_p,\ \ell_i,\ \kappa,\ \delta,\ \mathbf{g},\ \bm{I}_{PF}^{c},\ \mathbf{r}_X,\ \mathbf{z}_X]^T,
\end{eqnarray}
where $\ell_i$ is the normalized internal inductance, $\kappa$ and $\delta$ are
elongation and triangularity, $\mathbf{g}$ denotes plasma-wall gap metrics,
$\bm{I}_{PF}^{c}\in\mathbb{R}^{n}$ are the $n$ active PF (control) coil currents,
and $(\mathbf{r}_X,\mathbf{z}_X)$ are the upper and lower X-point coordinates.
For brevity, we separate the actuated control currents from the remaining, equilibrium-dependent
parameters by writing $\mathbf{s}=(\bm{\xi},\,\bm{I}_{PF}^{c})$, where
$\bm{\xi}\equiv[I_p,\ \ell_i,\ \kappa,\ \delta,\ \mathbf{g},\ \mathbf{r}_X,\ \mathbf{z}_X]^T$
collects the components of $\mathbf{s}$ other than $\bm{I}_{PF}^{c}$. So the ML surrogate (described in Sec.\ref{SEC.2.2} ) predicts,
\begin{eqnarray}
    \widehat{\gamma}_z(t) = f_\theta(\mathbf{s}(t)) = f_\theta\!\left(\bm{\xi}(t),\,\bm{I}_{PF}^{c}(t)\right).
\end{eqnarray}}

\textcolor{black}{The feedback requires the gradient of $\widehat{\gamma}_z$ with respect to the actuated
currents. The sensitivity row vector,
\begin{eqnarray}
    \mathbf{S}_\gamma(t) \equiv
    \left.\frac{\partial \widehat{\gamma}_z}{\partial \bm{I}_{PF}^{c}}\right|_{\bm{\xi}}
    \in \mathbb{R}^{1\times n}
\end{eqnarray}
is computed in real time via automatic differentiation and is evaluated at the current
operating point $\mathbf{s}(t)$. It is a \emph{partial} derivative with respect to the
control currents, taken with the remaining parameters $\bm{\xi}$ held fixed. Here
$\bm{I}_{PF}^{c}$ are the only actuated degrees of freedom, as distinct from the full
external conductor vector $\mathbf{I}_e=(\bm{I}_{PF}^{c},\,\mathbf{I}_{\mathrm{pass}})$
of Eq.~\eqref{eqn:2}, which also contains the passive structure currents.}

\textcolor{black}{The true plant response to a change in the coil currents is the \emph{total} derivative,
which adds to $\mathbf{S}_\gamma$ the indirect effect transmitted through the equilibrium
response of the dependent parameters:
\begin{eqnarray}
    \frac{d\gamma_z}{d\bm{I}_{PF}^{c}}
    = \underbrace{\left.\frac{\partial \gamma_z}{\partial \bm{I}_{PF}^{c}}\right|_{\bm{\xi}}}_{\displaystyle \mathbf{S}_\gamma}
    + \sum_{\xi_k\in\bm{\xi}}\frac{\partial \gamma_z}{\partial \xi_k}\,
      \frac{d \xi_k}{d \bm{I}_{PF}^{c}},
    \label{eq:total_vs_partial}
\end{eqnarray}
where the sum runs over the dependent parameters $\bm{\xi}$ only and \emph{not} over
$\bm{I}_{PF}^{c}$, so that the direct sensitivity appears once. Each factor
$d\xi_k/d\bm{I}_{PF}^{c}$ is the response of an equilibrium parameter (shape, internal
inductance, X-point geometry, or plasma-wall gap) to a change in the coil currents.}

\textcolor{black}{The passive structure currents $\mathbf{I}_{\mathrm{pass}}$ are dependent states rather
than inputs: they are not contained in $\mathbf{s}$, are induced by changes in
$\bm{I}_{PF}^{c}$ and by plasma motion, and are advanced time dependently by the circuit
dynamics of Eq.~\eqref{eq.3}. They therefore contribute neither a partial derivative
$\partial\gamma_z/\partial\mathbf{I}_{\mathrm{pass}}$ nor a separate term in
Eq.~\eqref{eq:total_vs_partial}. Their influence on the vertical mode is nonetheless fully
retained: it is carried by the plant model used for the closed-loop simulations, and it
enters the indirect channel of Eq.~\eqref{eq:total_vs_partial} implicitly, through the
equilibrium response $d\xi_k/d\bm{I}_{PF}^{c}$ of the parameters to the coils. The
controller uses only the partial derivative $\mathbf{S}_\gamma$ and therefore neglects this
cross-coupling sum.}

This approximation is justified on two grounds.  First,  within a single
control timestep $\Delta t = 1$~ms,  the equilibrium shift is small: at
moderate growth rates ($\gamma_{gr}\lesssim 6$~s$^{-1}$),  the
displacement amplification per cycle is $e^{\gamma_{gr}\Delta t}\lesssim
1.006$,  and the associated changes in $\ell_i$,  $\kappa$,  and gap
metrics are negligible relative to the direct $\bm{I}_{PF}^{c}$
effect.  Second,  PF5U/L,  the dominant stabilizing actuator,  modifies
the external field decay index primarily through a radial field
perturbation that couples weakly to the shape-setting quadrupole and
hexapole harmonics. \textcolor{black}{We do not rely on this argument alone: the finite-difference comparison reported in the following paragraph confirms that the neglected cross terms in Eq.~\eqref{eq:total_vs_partial} are small for this coil pair (median discrepancy $8\pm4\%$),  and larger for the PF coils.}
 
For coils (PF1 to 4),  whose fields simultaneously affect shape and
stability,  the cross terms may be non-negligible.  We quantified this
gap by computing the finite-difference total derivative
$\Delta\gamma_z/\Delta I^{c}_{\mathrm{PF}, k}$ from MEQ-FGE at 50 randomly
selected equilibria and comparing it to the surrogate partial derivative
$(\mathbf{S}_\gamma)_k$.  For PF5U/L the median relative discrepancy is
$8\pm 4\%$; for PF1 to 4 it rises to $18\pm 9\%$.  The QP regularization
term $w_V\|\mathbf{V}_{\mathrm{PF}}\|^2$ damps the effect of directional
errors on PF coils,  which partly explains why the controller succeeds
despite the larger partial derivative bias on those channels.  However, 
this sensitivity gap remains a source of systematic error that would
grow during active shape modification,  and is a primary motivation for
the cascade architecture discussed in Section~\ref{sec:shape_coupling}.
 
\textcolor{black}{At the ARC~V3A controllability boundary the single cycle amplification remains modest: even at $\gamma_{gr}=11.2$~s$^{-1}$ (case~A10) it is $e^{\gamma_{gr}\Delta t}\approx1.011$. The staleness that matters is therefore not single cycle but cumulative: over the ${\sim}89$ control cycles within one growth time at this rate (Section~\ref{sec:discussion}),  the linearization point drifts,  so $\bm{S}_\gamma$ becomes progressively misdirected relative to the evolving equilibrium. This accumulated linearization staleness,  compounding with voltage saturation,  contributes to the failures observed in cases A10 (highest growth rate, voltage saturated) and C6 (a large sawtooth crash like disturbance that saturates PF5 immediately) in Section~\ref{sec:results}.}

\textcolor{black}{The X-point locations are require both for the double-null flux-balance output $\psi_r$ and for the sensitivity row $\bm{S}_\psi$ introduced below.} We denote primary and secondary X-points by $X_p=(R_{X, p}, Z_{X, p})$ and $X_s=(R_{X, s}, Z_{X, s})$,  obtained from $\psi(R, Z)$ by solving $\nabla\psi\sim0$ in bounded search regions,  with the Newton iteration initialized from the previous cycle solution for robustness to noise.

Let $\bm{I}_{PF}^{c} \in \mathbb{R}^{n}$ be the vector of active PF coil currents,  $\mathbf{V}_{\mathrm{PF}}$ the applied voltages,  $\mathbf{L}_{\mathrm{PF}}$ the inductance matrix,  and $\mathbf{R}_{\mathrm{PF}}$ the resistance matrix. The circuit model is:
\begin{eqnarray}
    \mathbf{L}_{\mathrm{PF}}\frac{d\bm{I}_{PF}^{c}}{dt} + \mathbf{R}_{\mathrm{PF}}\bm{I}_{PF}^{c} = \mathbf{V}_{\mathrm{PF}} - \mathbf{M}_{\mathrm{PF}, p}\frac{d I_p}{dt}, 
\end{eqnarray}
where $\mathbf{M}_{\mathrm{PF}, p}$ captures mutual coupling from plasma evolution. A first order discretization over one control step $\Delta t$ yields:
\begin{multline}
    \bm{I}_{PF}^{c}(t+\Delta t) \approx \bm{I}_{PF}^{c}(t) \\
    + \Delta t \mathbf{L}_{\mathrm{PF}}^{-1} \left(\mathbf{V}_{\mathrm{PF}}(t) - \mathbf{R}_{\mathrm{PF}}\bm{I}_{PF}^{c}(t) - \mathbf{M}_{\mathrm{PF}, p}\dot I_p(t)\right).
    \label{eq:current_increment}
\end{multline}
We define the voltage-to-current incremental map $\mathbf{G}_{VI}\equiv\Delta t\, \mathbf{L}_{\mathrm{PF}}^{-1}$ and the known affine drift
\begin{eqnarray}
    \mathbf{d}(t) \equiv
      -\Delta t\, \mathbf{L}_{\mathrm{PF}}^{-1}
      \!\left(
        \mathbf{R}_{\mathrm{PF}}\bm{I}_{PF}^{c}(t)
        + \mathbf{M}_{\mathrm{PF}, p}\dot{I}_p(t)
      \right), 
    \label{eq:drift_term}
\end{eqnarray}
so that $\Delta\bm{I}_{PF}^{c} = \mathbf{G}_{VI}\mathbf{V}_{\mathrm{PF}}(t)+\mathbf{d}(t)$. During flattop ($|\dot{I}_p|<0.1$~MA/s) the plasma coupling contribution is small and $\Delta\bm{I}_{PF}^{c}$ is approximately affine in $\mathbf{V}_{\mathrm{PF}}$. We note that the FGEL state-space model
(Eq.~\ref{eq:current_increment}) includes a
$\delta\dot{\mathbf{u}}$ feed-through term \textcolor{black}{(recall $\delta\mathbf{u}$ denotes the PF coil voltage inputs)} through the matrix
$\mathbf{V} = -(\partial\mathbf{F}_D/\partial\dot{\mathbf{x}}_S)
(\partial\mathbf{x}_S/\partial\mathbf{u})$.  This term captures the
effect of voltage \emph{rate} on the dynamics.  In the controller,  the
QP optimises over $\mathbf{V}_{\mathrm{PF}}$ (voltage level),  not over
$\dot{\mathbf{V}}_{\mathrm{PF}}$.  The power supply model
(Eq.~\ref{eq:ps_model}) implicitly defines a voltage rate through the
first-order hold,  but this rate is not fed back into the FGEL-derived
prediction.  During steady-state operation where
$|\dot{\mathbf{V}}|$ is small,  the omission is benign.  During fast
transients, such as the initial convergence phase or disturbance
rejection, $\dot{\mathbf{V}}$ can be significant,  and neglecting the
$\mathbf{V}$ matrix contribution introduces a prediction error in
$\hat{\gamma}_z^+$.  As shown below,  the oscillatory behavior in case~C4
(Section~\ref{sec:results}) may partly originate from this unmodelled
voltage-rate coupling.  Incorporating $\dot{\mathbf{V}}_{\mathrm{PF}}$
into the QP as a decision variable or constraint is conceptually
straightforward but doubles the problem dimension; we defer this
extension to future work.

In double-null operation,  we regulate multiple quantities simultaneously:
\begin{eqnarray}
    \bm{y} \equiv [\gamma_{gr},  \psi_r,  z,  I_p]^T, 
    \label{eq:output_vector}
\end{eqnarray}
where $\psi_r \equiv \psi_{X, u} - \psi_{X, l}$ is the flux difference between upper and lower X-points. Linearizing with respect to the PF coil currents gives:
\begin{eqnarray}
    \delta \bm{y} \approx \bm{S}_y \,  \delta \bm{I}_{PF}^{c},  \quad \bm{S}_y \equiv \begin{bmatrix} \bm{S}_\gamma \\ \bm{S}_\psi \\ \bm{S}_z \\ \bm{S}_{I_p} \end{bmatrix}.
    \label{eq:output_sensitivity}
\end{eqnarray}
The rows of $\bm{S}_y$ are obtained from different sources,  reflecting
the mixed observability of the controlled outputs.  The growth rate
sensitivity $\bm{S}_\gamma = \partial\hat{\gamma}_z/\partial
\bm{I}_{PF}^{c}$ is computed via automatic differentiation of
the ML surrogate (Section~\ref{sec:physics}).  The remaining rows are
derived from the FGEL linearization at the current operating point:
$\bm{S}_\psi \equiv \partial\psi_r/\partial\bm{I}_{PF}^{c}$ is
extracted from the flux Jacobian $\partial\psi/\partial\mathbf{I}_e$
evaluated at the upper and lower X-point locations;
$\bm{S}_z \equiv \partial z/\partial\bm{I}_{PF}^{c}$ is
obtained from the current-centroid response in the FGEL vertical
eigenmode; and $\bm{S}_{I_p} \equiv
\partial I_p/\partial\bm{I}_{PF}^{c}$ is taken from the
plasma-current constraint row of the linearized circuit equations.
Because $\bm{S}_\gamma$ originates from the surrogate while
$\bm{S}_\psi$,  $\bm{S}_z$,  $\bm{S}_{I_p}$ originate from FGEL,  the
composite $\bm{S}_y$ mixes two model sources.  Consistency is
maintained to the extent that the surrogate was trained on FGEL
outputs; however,  residual inconsistencies between the surrogate
gradient and the FGEL Jacobian can cause the QP to allocate voltage
sub-optimally when trade-offs between $\gamma_{gr}$ and other outputs
are tight.

The $\bm{S}_{I_p}$ row is near-zero during flattop but retained for generality during current ramps. Combining with circuit dynamics:
\begin{eqnarray}
    \bm{y}(t+\Delta t) \approx \bar{\bm{y}}(t) + \bm{H}\bm{V}_{PF}(t), 
    \label{eq:output_prediction_compact}
\end{eqnarray}
where $\bm{H} \equiv \bm{S}_y \bm{G}_{VI}$ and $\bar{\bm{y}}(t) \equiv \bm{y}(t) + \bm{S}_y \bm{d}(t)$ is the zero-input prediction.

The coil voltage commands are computed by solving a constrained quadratic program at each control step:
\begin{eqnarray}
    \min_{\mathbf{V}_{\mathrm{PF}}}\ \ J(\mathbf{V}_{\mathrm{PF}}), 
\end{eqnarray}
with cost function
\begin{align}
    J &\equiv w_\gamma\left(\widehat{\gamma}_z^+-\gamma_{z, \mathrm{tgt}}\right)^2
      + w_{\psi}\left(\psi_r^+-\psi_{r, \mathrm{tgt}}\right)^2 \nonumber \\
      &+ w_z\left(z^+-z_{\mathrm{ref}}\right)^2
      + w_I\left(I_p^+-I_{p, \mathrm{ref}}\right)^2
      + w_V\|\mathbf{V}_{\mathrm{PF}}\|_2^2, 
\end{align}
where superscript $+$ denotes predicted values at $t+\Delta t$,  subject to
\begin{align}
\bm{V}_{\min} \le \bm{V}_{\mathrm{PF}} \le \bm{V}_{\max},  \label{eq:Vlims}\\
\bm{I}_{\min} \le \bm{I}_{PF}^{c}(t+\Delta t) \le \bm{I}_{\max}. \label{eq:Ilims}
\end{align}
\textcolor{black}{Here $\gamma_{z, \mathrm{tgt}}$ is the target growth rate setpoint. For the elongated double-null ARC~V3A plasma the $n=0$ vertical mode is open-loop unstable ($\gamma_{gr}>0$),  and out-vessel coils alone cannot render it passively stable ($\gamma_{gr}<0$). The controller therefore does not target passive decay; it regulates $\gamma_{gr}$ to a small \emph{positive} setpoint (here $5$~s$^{-1}$,  Section~\ref{sec:results}) chosen so that the residual instability is slow enough to remain within the correction bandwidth of the available actuators.} We take $\psi_{r, \mathrm{tgt}}=0$~Wb for balanced DN,  and the weights encode both control priorities and dimensional normalization. For ARC~V3A: $w_\gamma = 60$ (increased to 200 when $\gamma_{gr} > 20$~s$^{-1}$),  $w_\psi = 50$,  $w_z = 30$,  $w_{I_p} = 10$,  and $w_V = 0.01$. \textcolor{black}{The weights combine dimensional normalization with control priority. Each output is first scaled by its characteristic magnitude (s$^{-1}$ for $\gamma_{gr}$,  Wb for $\psi_r$,  m for $z$,  A for $I_p$) so the tracking terms are comparable; the normalized terms are then order by priority,  with growth rate regulation weighted highest and a small $w_V$ penalizing actuator effort without dominating tracking. The growth rate weight is raised to $w_\gamma=200$ when $\gamma_{gr}>20$~s$^{-1}$ to prioritize stabilization under fast-growth transients. The values quoted were tuned on the baseline case (A1) and held fixed across the full test matrix.}

We note that the QP cost does not include explicit boundary shape terms. The controlled outputs $\gamma_{gr}$,  $\psi_r$,  $z$,  and $I_p$ constrain the equilibrium but do not uniquely determine the separatrix geometry. The mapping from shape parameters $(\kappa,  \delta,  g_{\rm in},  g_{\rm out})$ to $\gamma_{gr}$ is many-to-one: distinct equilibria can share the same growth rate while differing in elongation,  triangularity, or plasma-wall gaps. Consequently, the QP does not prevent slow shape drift on timescales of seconds if the $\gamma_{gr}$ optimal coil current trajectory departs from the shape-maintaining trajectory. In the present simulations,  shape is sustained by the feedforward scenario design. A separate shape control layer either as additional terms in the QP or as an outer-loop controller operating on longer timescales would be required for sustained scenario evolution or active shape modification during $\gamma_{gr}$ regulation. This point is discussed further in Section~\ref{sec:shape_coupling}.

Substituting the affine prediction~\eqref{eq:output_prediction_compact} into the cost yields a quadratic objective:
\begin{eqnarray}
    J = \bm{V}_{PF}^T \bm{Q} \bm{V}_{PF} + \bm{q}^T \bm{V}_{PF} + \mathrm{const}, 
    \label{eq:qp_quadratic_form}
\end{eqnarray}
where
\begin{eqnarray}
    \bm{Q} = w_\gamma \bm{H}_\gamma^T \bm{H}_\gamma + w_\psi \bm{H}_\psi^T \bm{H}_\psi + w_z \bm{H}_z^T \bm{H}_z \nonumber \\ + w_I \bm{H}_{I_p}^T \bm{H}_{I_p} + w_V \bm{I}.
    \label{eq:Q_matrix}
\end{eqnarray}
Since each $\bm{H}_i^T \bm{H}_i$ is positive semidefinite and $w_V \bm{I}$ is positive definite (for $w_V > 0$),  the Hessian $\bm{Q}$ is positive definite,  guaranteeing a unique global minimum. The constraints are linear in $\bm{V}_{PF}$,  so this is a strictly convex QP solvable via OSQP~\cite{Stellato2020}.

The applied voltage is modeled with a first-order hold:
\begin{eqnarray}
    V^{\rm appl}_{k+1}=a_v\,  V^{\rm appl}_k+(1-a_v)\, V^{\rm cmd}_k, 
    \label{eq:ps_model}
\end{eqnarray}
where $a_v=e^{-\Delta t/\tau_{\rm PS}}$ and $\tau_{\rm PS}=0.5$~ms is the baseline power supply time constant ($\approx 320$~Hz bandwidth). Equilibrium reconstruction is available with a delay of $d$ samples,  so the QP uses the most recent $\bm{y}_{k-d}$ and $\bm{S}_{y, k-d}$; PF currents are propagated forward using the measured voltage history.

The algorithm executes at each cycle:
\begin{enumerate}[noitemsep]
    \item Reconstruct equilibrium and assemble $\mathbf{s}(t)$;
    \item Evaluate $\widehat{\gamma}_z(t)=f_\theta(\mathbf{s}(t))$ and compute $\mathbf{S}_\gamma(t)$;
    \item Update $\mathbf{S}_y(t)$ and $\mathbf{G}_{VI}$;
    \item Solve the QP for $\mathbf{V}_{\mathrm{PF}}(t)$;
    \item Apply voltages and repeat.
\end{enumerate}
This is a one step receding horizon controller re-linearized each cycle through $\bm{S}_{y, k}$; it is not a PID formulation,  and integral action is not implicit. If steady state offset elimination under model mismatch is required,  an explicit integral augmentation can be added.

\section{Closed-Loop Simulation Results}\label{sec:results}

\subsection{Simulation environment}

Closed-loop performance is evaluated using an offline simulation environment that couples the MEQ free-boundary equilibrium evolution code to PF circuit and power supply dynamics. The equilibrium solver advances the Grad-Shafranov system with passive structure eddy currents and plasma wall coupling computed self-consistently at a 1~ms sample time. Diagnostic signals are extracted directly from the equilibrium solver; no measurement noise or reconstruction latency is imposed unless explicitly stated (see Groups~B and C below and Section~\ref{sec:discussion}).

Simulations assume quasi-static evolution on the resistive timescale,  consistent with flattop phases where $|\dot{I}_p| < 0.1$~MA/s. During rapid current ramps or fast MHD transients ($|\dot{I}_p| > 0.5$~MA/s),  this assumption may break down and a high bandwidth fallback controller would be required.

The power supply time constants ($\tau_{\rm PS}$), voltage limits ($V_{\max}$), and coil circuit parameters ($\mathbf{L}_{\mathrm{PF}}$, $\mathbf{R}_{\mathrm{PF}}$) used throughout this work are representative values chosen to exercise the model and to span a plausible actuator envelope; they are intended to demonstrate the control framework and do not correspond to engineering design values for ARC~V3A. The reported pass/marginal/fail outcomes should accordingly be read as a map of where this class of controller and actuation is and is not effective, not as engineering acceptance results for any specific hardware.

\subsection{Test matrix}

To assess the controller across the operating space rather than at a single equilibrium,  we constructed a 24-case simulation matrix organized in three groups. Group~A varies equilibrium conditions at fixed actuator settings. Group~B tests actuator and diagnostic degradation at the baseline equilibrium. Group~C applies transient disturbances during steady-state operation.

Control references are $\gamma_{z, \mathrm{tgt}}=5$~s$^{-1}$ (see
Section~\ref{sec:controller} for the rationale of a positive target), 
$I_{p, \mathrm{ref}}=12.4$~MA, 
$z_{\mathrm{ref}}\sim 2\times10^{-5}$~m, 
$\psi_{r, \mathrm{tgt}}\sim 0$~Wb (symmetric DN). QP weights are as specified in Section~\ref{sec:controller}. Separatrix elongation is not imposed in the QP cost; the observed $\kappa_{\rm sep}$ values reported below are emergent outcomes of the equilibrium response to the controller's voltage commands,  constrained by the feedforward coil current trajectory. Whether these values would be sustained over longer simulation windows or during active shape changes is not tested here (see Section~\ref{sec:shape_coupling}).

Target criteria for a ``Pass'' outcome are: convergence time $\tau_{\mathrm{conv}}<2$~s,  $\mathrm{RMS}(\gamma_{gr}-\gamma_{z, \mathrm{tgt}})<0.3$~s$^{-1}$,  $|\Delta Z_{\max}|<5\%\, a$,  and $|\psi_r|<0.03$~Wb. These thresholds are diagnostic targets for assessing controllability,  not engineering acceptance requirements; they were chosen as follows. The convergence bound $\tau_{\mathrm{conv}}<2$~s is a small fraction of the $\sim$15-min flat-top and is comfortably faster than the second-scale timescale on which shape and current profiles evolve. The growth rate bound of $0.3$~s$^{-1}$ is $\sim$7\% of the baseline open-loop growth rate ($4.2$~s$^{-1}$), small enough that the regulated mode stays well inside the actuator correction bandwidth. The vertical-excursion bound $|\Delta Z_{\max}|<5\%\,a$ keeps the plasma clear of the first wall and within the range where the linearized response remains valid. The flux-imbalance bound $|\psi_r|<0.03$~Wb preserves the up/down symmetry of the double-null and avoids drift toward a single-null configuration. ``Marginal'' indicates criteria are met narrowly or one metric is intermittently violated. ``Fail'' indicates sustained violation or loss of control.

\subsection{Group A: Equilibrium variation}

\begin{table*}[ht]
\centering
\caption{Group~A: Closed-loop performance under equilibrium
variation. All cases use $V_{\max}=500$~V, zero diagnostic
latency, and baseline power-supply bandwidth
($\tau_{\mathrm{PS}}=0.5$~ms). $\gamma_z(t_0)$ is the growth rate of the
initial equilibrium before the controller acts. ``Outcome'' indicates whether the
controller meets all design criteria
($\tau_{\mathrm{conv}}<2$~s,
$\mathrm{RMS}(\gamma_z-\gamma_{z,\mathrm{tgt}})<0.3$~s$^{-1}$,
$|\Delta Z_{\max}|<5\%\,a$, $|\psi_r|<0.03$~Wb).}
\label{tab:groupA}
\small
\begin{tabular}{clcccccccc}
\toprule
Case
  & Description
  & $\kappa_{\mathrm{sep}}$
  & $\ell_i$
  & $\beta_p$
  & $\gamma_z(t_0)$ [s$^{-1}$]
  & $\tau_{\mathrm{conv}}$ [s]
  & RMS $\gamma_z$ err [s$^{-1}$]
  & $|\Delta Z_{\max}|/a$ [\%]
  & Outcome \\
\midrule
A1
  & Baseline
  & 1.80 & 0.65 & 0.524
  & $4.2$
  & 0.5
  & 0.08
  & 1.8
  & \textbf{Pass} \\
A2
  & High-$\kappa_{sep}$
  & 1.90 & 0.65 & 0.52
  & $7.8$
  & 1.2
  & 0.24
  & 3.9
  & \textbf{Marginal} \\
A3
  & Moderate $\kappa_{sep}$
  & 1.70 & 0.65 & 0.52
  & $1.8$
  & 0.3
  & 0.04
  & 0.6
  & \textbf{Pass} \\
A4
  & Peaked $\ell_i$
  & 1.80 & 0.90 & 0.52
  & $6.1$
  & 0.9
  & 0.19
  & 2.8
  & \textbf{Pass} \\
A5
  & Broad $\ell_i$
  & 1.80 & 0.50 & 0.52
  & $3.1$
  & 0.4
  & 0.06
  & 1.2
  & \textbf{Pass} \\
A6
  & Low $\beta_p$
  & 1.80 & 0.65 & 0.5
  & $5.8$
  & 0.8
  & 0.16
  & 2.5
  & \textbf{Pass} \\
A7
  & $\Delta \beta_{p} =5$\%
  & 1.80 & 0.90 & 0.49 
  & $8.9$
  & 1.8
  & 0.35
  & 4.6
  & \textbf{Marginal}$^\dagger$ \\
A8
  & Outer gap (1~cm)
  & 1.80 & 0.65 & 0.52
  & $3.5$
  & 0.4
  & 0.07
  & 1.4
  & \textbf{Pass} \\
A9
  & Outer gap (3.5~cm)
  & 1.80 & 0.65 & 0.52
  & $5.4$
  & 0.7
  & 0.13
  & 2.2
  & \textbf{Pass} \\
A10
  & High $\kappa$ + peaked $\ell_i$
  & 1.90 & 0.90 & 0.60
  & $11.2$
  & ---
  & ---
  & $>8$
  & \textbf{Fail} \\
\bottomrule
\multicolumn{10}{l}{\footnotesize $^\dagger$A7 meets
criteria narrowly; $\gamma_z$ intermittently exceeds target
band during transient ($t<2$~s). Steady-state regulation}\\
\multicolumn{10}{l}{\footnotesize achieved but with
persistent offset of ${\sim}0.2$~s$^{-1}$ from target. A10
diverges within 0.8~s; the open-loop growth rate exceeds}\\
\multicolumn{10}{l}{\footnotesize the controller bandwidth
with 500~V voltage authority.}
\end{tabular}
\end{table*}

The baseline case (A1,  $\kappa_{\mathrm{sep}}=1.80$,  $\ell_i=0.65$,  $\beta_p=0.60$) converges from $\gamma_{gr}(t_0)=+4.2$~s$^{-1}$ to the target within 0.5~s with steady-state RMS error of 0.08~s$^{-1}$. Plasma current remains within $12\pm 0.05$~MA (peak),  vertical excursion stays below 2\% of the minor radius,  and DN flux balance is maintained at $|\psi_r|<0.015$~Wb. The separatrix elongation is sustained at $1.80\pm 0.02$.

\begin{figure}[ht!]
\centering
\includegraphics[width=\columnwidth]{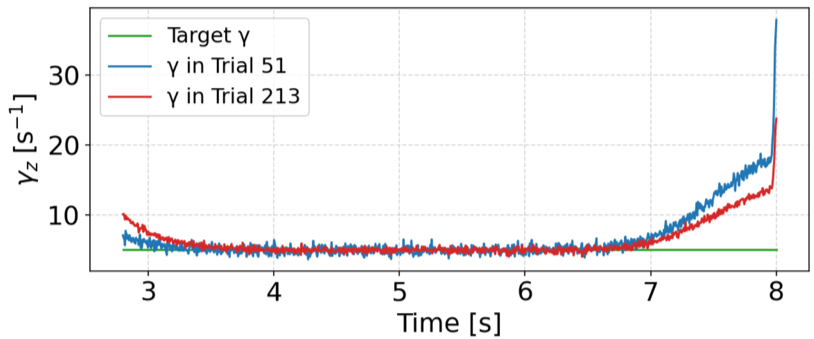}\\
\includegraphics[width=\columnwidth]{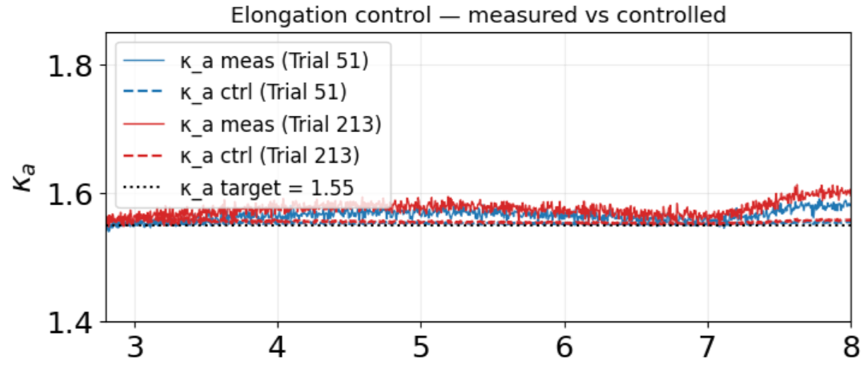}\\
\includegraphics[width=\columnwidth]{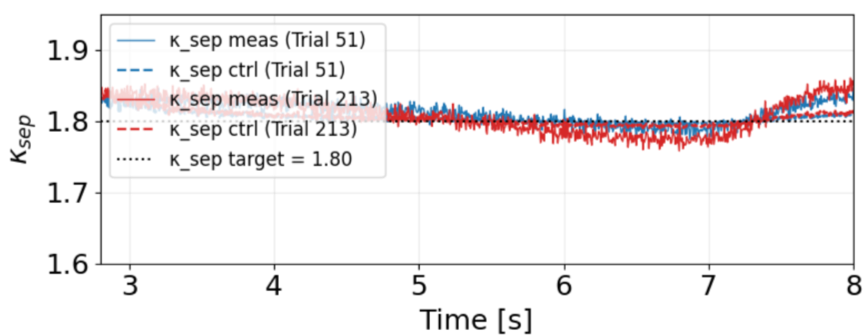} \\
\includegraphics[width=\columnwidth]{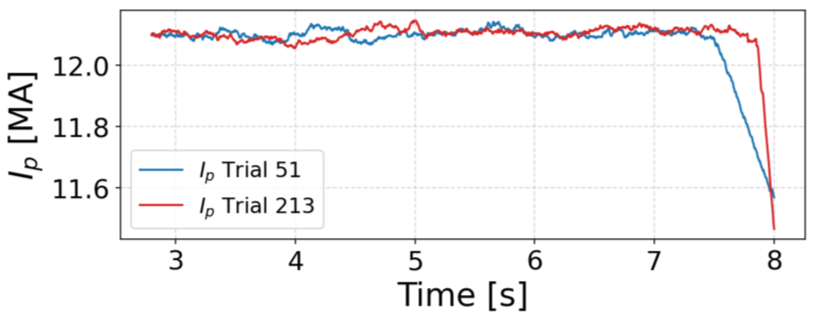} \\
\includegraphics[width=\columnwidth]{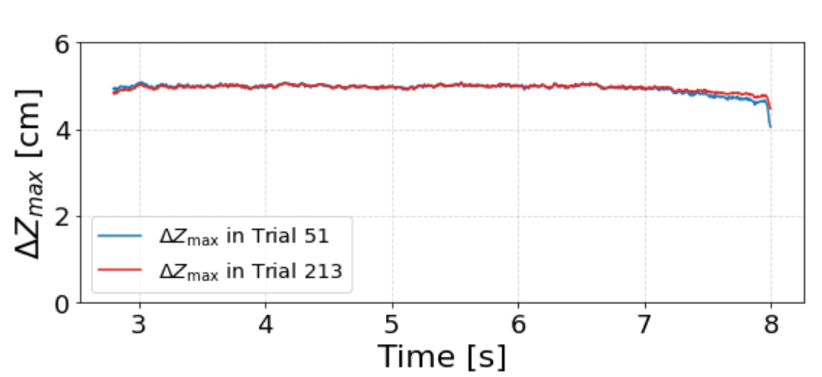}
\caption{Time traces during ARC~V3A flattop regulation (baseline case A1,  $t=3$ to $8$~s). From top to bottom: vertical growth rate $\gamma_{gr}$,  areal elongation $\kappa_{\mathrm{areal}}$,  separatrix elongation $\kappa_{\mathrm{sep}}$,  plasma current $I_p$,  and maximum vertical displacement $\Delta Z_{\max}$.}
\label{fig:arc_trial51}
\end{figure}

\textcolor{black}{Figure~\ref{fig:arc_trial51} shows the baseline (A1) closed-loop response over the $t=3$ to $8$~s flat-top window,  with one panel per controlled or diagnostic quantity. The top panel shows $\gamma_{gr}$ driven from its initial $+4.2$~s$^{-1}$ to the target and held there; the areal- and separatrix-elongation panels show that $\kappa_{\mathrm{areal}}$ and $\kappa_{\mathrm{sep}}$ remain within $\pm0.02$ of their nominal values,  confirming that the growth rate regulation does not drive a shape excursion over this window (Section~\ref{sec:shape_coupling}); the $I_p$ panel shows the plasma current held within $12\pm0.05$~MA; and the $\Delta Z_{\max}$ panel shows the vertical excursion bounded below 2\% of the minor radius. The rise in $\gamma_{gr}$ and the departure of $I_p$ near $t\approx8$~s mark the end of the controlled flat-top window and are not a controller-induced disruption.}

Moderate elongation (A3,  $\kappa_{\mathrm{sep}}=1.70$) converges fastest and with the smallest errors,  as expected from the lower growth rate. Cases A4 to A6 isolate the sensitivity to $\ell_i$ and $\beta_p$ individually: peaked current profiles (A4) raise the initial growth rate to $+6.1$~s$^{-1}$ and roughly double the steady state tracking error relative to baseline,  while low $\beta_p$ (A6) produces comparable degradation.

Case A7 (combined deteriorated conditions: $\ell_i=0.90$,  $\beta_p=0.10$) is marginal. The initial growth rate of $+8.9$~s$^{-1}$ forces PF5 voltage saturation during the first 1.5~s,  and the controller achieves steady-state regulation only with a persistent offset of ${\sim}0.2$~s$^{-1}$ from target. This is the direct analog of Nishio et~al.'s deteriorated plasma finding~\cite{nishio1993} for the ARC geometry: the combination of low pressure and peaked current drives the equilibrium toward the controllability boundary.

Case A10 ($\kappa_{\mathrm{sep}}=1.90$,  $\ell_i=0.90$) fails. The initial growth rate ($+11.2$~s$^{-1}$) exceeds what the 500~V voltage authority can counteract through out-vessel coils. Vertical displacement exceeds 8\% of the minor radius within 0.8~s. This defines the upper boundary of the controllable operating space for this actuator configuration.



\subsection{Group B: Actuator and diagnostic degradation}

\begin{table*}[ht]
\centering
\caption{Group~B: Performance degradation under actuator and
diagnostic limitations. All cases use the baseline equilibrium
(A1: $\kappa_{\mathrm{sep}}=1.80$, $\ell_i=0.65$,
$\beta_p=0.524$, $\gamma_z(t_0)=+4.2$~s$^{-1}$)
unless noted. Design criteria as in Table~\ref{tab:groupA}.
Parenthetical entries for B6 indicate the residual actuator
set after failure.}
\label{tab:groupB}
\small
\begin{tabular}{clccccccc}
\toprule
Case
  & Description
  & $V_{\max}$ [V]
  & $\tau_{\mathrm{PS}}$ [ms]
  & Latency [ms]
  & $\tau_{\mathrm{conv}}$ [s]
  & RMS $\gamma_z$ [s$^{-1}$]
  & $|\Delta Z_{\max}|/a$ [\%]
  & Outcome \\
\midrule
B1
  & Reduced $V_{\max}$
  & 200 & 0.5 & 0
  & 1.4
  & 0.21
  & 3.2
  & \textbf{Marginal} \\
B2
  & Severely reduced $V_{\max}$
  & 100 & 0.5 & 0
  &  
  &  
  & $>6$
  & \textbf{Fail} \\
B3
  & Slow PS ($\tau_{\rm PS}=2$~ms)
  & 500 & 2.0 & 0
  & 0.9
  & 0.14
  & 2.6
  & \textbf{Pass} \\
B4
  & 1~ms latency
  & 500 & 0.5 & 1
  & 0.7
  & 0.12
  & 2.3
  & \textbf{Pass} \\
B5
  & 3~ms latency
  & 500 & 0.5 & 3
  & 1.6
  & 0.28
  & 4.1
  & \textbf{Marginal} \\
B6
  & PF5 pair failure (one coil)
  & 500 & 0.5 & 0
  & 2.1
  & 0.38
  & 4.8
  & \textbf{Marginal}$^\dagger$ \\
\midrule
B7
  & Low $V_{\max}$ + 3~ms latency
  & 200 & 0.5 & 1
  &  
  &  
  & $>5$
  & \textbf{Fail} \\
B8
  & Slow PS + 8~ms latency
  & 500 & 2.0 & 3
  &  
  &  
  & $>7$
  & \textbf{Fail} \\
\bottomrule
\multicolumn{9}{l}{\footnotesize $^\dagger$B6: QP reallocates
to PF3/4 but coupling is weaker; steady-state shows
${\sim}0.3$~s$^{-1}$ persistent $\gamma_z$ offset and}\\
\multicolumn{9}{l}{\footnotesize degraded DN balance
($|\psi_r|\approx 0.04$~Wb, exceeding specification). B2
saturates PF5 voltage within the first}\\
\multicolumn{9}{l}{\footnotesize 50~ms and vertical
displacement grows monotonically. B7 and B8 show that
combining two individual degradations that}\\
\multicolumn{9}{l}{\footnotesize are each marginal/passing
alone produces failure a finding consistent with
Nelson~et~al.~\cite{nelson2024_sparc_vs}.}
\end{tabular}
\end{table*}

reducing the voltage limit to 200~V (B1) is marginal: the controller recovers but with degraded convergence time and tracking accuracy. At 60~V (B2) the controller fails; the current slew rate through the out-vessel coils is insufficient to counteract the baseline $+4.2$~s$^{-1}$ growth rate.

Diagnostic latency of 1~ms (B4) is well tolerated,  with modest performance degradation. At 3~ms (B5) performance becomes marginal. The characteristic growth time $\tau_\gamma = 1/4.2 \approx 238$~ms provides substantial margin relative to the 1~ms control cycle,  so the degradation is not primarily a raw timing issue. Its rather,  with 3~ms latency the QP operates on sensitivity information that is three cycles stale. Over the hundreds of control cycles within one growth time,  the accumulated directional error in $\bm{S}_\gamma$ degrades voltage allocation accuracy,  producing oscillatory tracking and increased steady-state error. This accumulated-staleness mechanism is distinct from the single-cycle amplification argument that governs fast-growing instabilities on existing devices.

PF5 single coil failure (B6) is marginal: the QP reallocates to PF3/4 but the weaker coupling produces a persistent $\gamma_{gr}$ offset and degraded DN balance ($|\psi_r|\approx 0.04$~Wb).

The most important finding in Group~B is that combining two individually tolerable degradations produces failure. Cases B7 (200~V + 1~ms latency) and B8 (slow PS + 3~ms latency) both fail,  despite each constituent degradation being manageable alone. This compound-degradation sensitivity is consistent with findings by Nelson et~al.~\cite{nelson2024_sparc_vs} for SPARC,  where the percentage of equilibria successfully caught dropped sharply when both voltage limits and filter latency increased together.

\subsection{Group C: Adding Disturbances}
\begin{table*}[ht]
\centering
\caption{Group~C: Disturbance rejection performance. All
disturbances applied at $t=5$~s during steady-state operation
from the baseline equilibrium (A1). ``Recovery time'' is time
to return $\gamma_z$ within $\pm 3$~s$^{-1}$ of target after
the disturbance. ``Peak excursion'' is the maximum
$|\Delta Z|/a$ during the transient.}
\label{tab:groupC}
\small
\begin{tabular}{clccccc}
\toprule
Case
  & Disturbance
  & Recovery [s]
  & Peak $|\Delta Z|/a$ [\%]
  & Peak $|\psi_r|$ [Wb]
  & Peak $|V_{\mathrm{PF5}}|$ [V]
  & Outcome \\
\midrule
C1
  & Step $\Delta\ell_i = +0.10$ 
  & 0.4
  & 2.1
  & 0.018
  & 320
  & \textbf{Pass} \\
C2
  & Step $\Delta\beta_p = -0.20$ 
  & 0.3
  & 1.5
  & 0.012
  & 260
  & \textbf{Pass} \\
C3
  & $\Delta I_{PF3}$ ($+5$~kA step)
  & 0.2
  & 0.8
  & 0.022
  & 180
  & \textbf{Pass} \\
C4
  &  $\Delta\ell_i\!=\!+0.20$,
    $\Delta\beta_p\!=\!-0.30$
  & 1.4
  & 4.3
  & 0.038
  & 480$^\dagger$
  & \textbf{Marginal}$^\ddagger$ \\
C5
  & Slow $\kappa$ ramp $1.70\to 1.90$ over 2~s
  & --$^\S$
  & 3.1
  & 0.025
  & 350
  & {Marginal} \\
C6
  &  $\Delta\ell_i = +0.25$ (monster sawtooth)
  & --
  & $>6$
  & $>0.05$
  & 500 (sat.)
  & \textbf{Fail} \\
\bottomrule
\multicolumn{7}{l}{\footnotesize $^\dagger$Near voltage
saturation (500~V limit); controller recovers but with
sustained oscillation for ${\sim}0.5$~s.}\\
\multicolumn{7}{l}{\footnotesize $^\ddagger$C4: The combined
disturbance pushes the equilibrium outside the surrogate
training envelope. $\hat{\gamma}_z$ prediction}\\
\multicolumn{7}{l}{\footnotesize error increases to
${\sim}0.4$~s$^{-1}$ during the transient, degrading
sensitivity accuracy. Controller recovers once the}\\
\multicolumn{7}{l}{\footnotesize equilibrium drifts back
toward the training distribution under closed-loop
regulation.}\\
\multicolumn{7}{l}{\footnotesize $^\S$C5: $\gamma_z$
tracking degrades continuously as $\kappa$ increases;
controller maintains stability but RMS error grows}\\
\multicolumn{7}{l}{\footnotesize from 0.08 to
0.32~s$^{-1}$ as $\kappa_{\mathrm{sep}}$ passes through 1.85.
At $\kappa_{\mathrm{sep}}=1.90$ the operating point}\\
\multicolumn{7}{l}{\footnotesize approaches the edge of
the training distribution and sensitivity estimates become
unreliable.}\\
\multicolumn{7}{l}{\footnotesize C6: PF5 voltage saturates
immediately; vertical displacement exceeds 6\% of minor
radius within 0.4~s. This represents}\\
\multicolumn{7}{l}{\footnotesize a disturbance beyond the
controller's allocation with out-vessel actuation at 500~V.}
\end{tabular}
\end{table*}

Single disturbances (C1 to C3) are well rejected. A sawtooth-like $\ell_i$ perturbation of $+0.10$ (C1) is corrected within 0.4~s; a confinement drop of $\Delta\beta_p = -0.20$ (C2) within 0.3~s; a spurious PF3 current perturbation (C3) within 0.2~s.

The combined disturbance C4 ($\Delta\ell_i=+0.20$,  $\Delta\beta_p=-0.30$) is marginal. The equilibrium moves outside the surrogate training envelope,  increasing the $\hat{\gamma}_z$ prediction error to ${\sim}0.4$~s$^{-1}$ during the transient. PF5 voltage approaches saturation (480~V),  and the controller exhibits sustained oscillation for ${\sim}0.5$~s before recovering. The sensitivity vector $\bm{S}_\gamma$ retains approximately correct direction but incorrect magnitude,  producing overshoot.

Case C6 ($\Delta\ell_i = +0.25$,  an extreme sawtooth-crash-like internal-inductance jump) fails: the disturbance is large enough that PF5 saturates immediately,  before the QP can redistribute effort,  and vertical displacement exceeds 6\% of the minor radius within 0.4~s. This disturbance exceeds the controller's authority with 500~V out-vessel actuation.

Case C5 (slow $\kappa$ ramp from 1.70 to 1.90 over 2~s) is marginal: $\gamma_{gr}$ tracking degrades continuously as $\kappa$ increases,  with RMS error growing from 0.08 to 0.32~s$^{-1}$ as $\kappa_{\mathrm{sep}}$ passes through 1.85. At $\kappa_{\mathrm{sep}}=1.90$ the operating point approaches the edge of the training distribution and sensitivity estimates become unreliable.

In summary,  the controller passes 12,  achieves marginal performance in 8,  and fails in 4. Failures occur at high growth rates with limited voltage authority (A10),  under combined actuator-diagnostic degradations (B7,  B8),  and for large transient disturbances that exceed actuator bandwidth or push the equilibrium beyond the surrogate training distribution (C6).

\section{Discussion}\label{sec:discussion}

Three mechanisms account for the observed performance: first,  growth rate feedback provides a stability-margin signal before large vertical motion develops. Position-only feedback requires a detectable displacement before correction begins; since $\delta z(t)\approx\delta z_0 e^{\gamma_{gr} t}$,  reducing $\gamma_{gr}$ directly extends the time window for correction. In the ARC~V3A baseline,  $\tau_\gamma = 1/4.2 \approx 238$~ms,  providing approximately 238 control cycles per growth time at 1~kHz. This margin is substantially more favourable than the fastest scenarios considered by Nelson et~al.~\cite{nelson2024_sparc_vs} for SPARC,  where growth rates up to ${\sim}60$~s$^{-1}$ compress the available correction window to ${\sim}17$~ms. Even in the most challenging ARC~V3A case (A10,  $\gamma_{gr} = 11.2$~s$^{-1}$,  $\tau_\gamma \approx 89$~ms),  the controller has approximately 89 cycles per growth time,  which remains adequate in principle but becomes insufficient when voltage saturation prevents the commanded corrections from being applied.

Second,  the sensitivity vector $\mathbf{S}_\gamma(t)=\partial\hat{\gamma}_z/\partial\bm{I}_{PF}^{c}$,  recomputed each cycle via automatic differentiation,  directs voltage to whichever coils have the highest leverage at that instant. In all passing cases PF5U/L carried the dominant stabilizing effort,  consistent with the strong dependence on coil-plasma distance,  while PF coils contributed mainly to shape control. Fixed gain controllers cannot adapt when actuator effectiveness shifts a limitation noted by Pesamosca et~al.~\cite{pesamosca2022_tcv} for TCV,  where even $H_\infty$ designs require gain scheduling across elongation ranges.

Third,  the QP resolves the coupling inherent to DN operation,  where a single PF circuit simultaneously affects stability margin,  flux balance,  vertical position,  and plasma current. Classical separate PD loops manage these conflicts through iterative hand-tuning; the QP encodes priorities through explicit weights and finds the constrained trade-off at each cycle. Nishio et~al.~\cite{nishio1993} recognized early that PD control with voltage limitation outperforms unconstrained PD by applying lower voltage for longer duration rather than a large initial impulse; the QP generalizes this insight to multivariable operation with explicit actuator constraints.

Supplying $P'(\psi_N)$ and $TT'(\psi_N)$ as explicit surrogate inputs,  rather than relying solely on $\ell_i$ and $\beta_N$,  is motivated by the sensitivity of the non-rigid vertical mode to perturbed current density structure \cite{Kumar_2026}. The profile aware model achieves MAE of 0.18~s$^{-1}$ versus 0.29~s$^{-1}$ for the scalar only model (38\% degradation). The improvement concentrates in equilibria with steep pedestal gradients and non-monotonic $TT'$ profiles.

\subsection{Coupling between \texorpdfstring{$\gamma_{gr}$}{gamma\_z} regulation and boundary shape}\label{sec:shape_coupling}

The QP cost function regulates $\gamma_{gr}$,  $\psi_r$,  $z$,  and $I_p$,  but does not include explicit boundary shape terms such as elongation,  triangularity,  or plasma-wall gaps. This omission reflects a deliberate scope limitation: the present work addresses vertical stability control,  not integrated shape regulation. However,  the coupling between $\gamma_{gr}$ actuation and shape evolution deserves discussion,  as it has direct implications for how this controller would interface with a broader control architecture.

The vertical growth rate depends on the external field decay index $n$,  the inductive coupling between the plasma current distribution and the passive structures,  and the non-rigid plasma response governed by $P'$ and $TT'$. All of these quantities are functions of the boundary shape. The mapping from shape parameters $(\kappa,  \delta,  \ell_i,  \beta_p,  g)$ to $\gamma_{gr}$ is many-to-one: multiple equilibria with distinct separatrix geometries can share the same growth rate. Regulating $\gamma_{gr}$ to a target value constrains the plasma to a level set in shape space a hypersurface of constant $\gamma_{gr}$ but does not determine the location on that surface.

In the baseline case (A1),  the controller maintains $\kappa_{\rm sep} = 1.80 \pm 0.02$ over the 5~s simulation window. This stability is not guaranteed by the control architecture; rather,  it reflects two features of the simulation setup. First,  PF5U/L carries the dominant $\gamma_{gr}$ regulation effort through differential mode commands that primarily produce a radial field perturbation,  modifying the effective decay index without strongly changing the quadrupole and hexapole field components that set elongation and triangularity. Second,  the 5~s simulation window is short relative to the timescale on which cumulative PF current drifts would modify the equilibrium shape.

A concrete conflict arises during active shape modification. If $\kappa_{\rm sep}$ were ramped upward (as in case C5),  the associated increase in $\gamma_{gr}$ would trigger corrective voltages from the QP that partially oppose the field changes needed to achieve higher elongation. The $\gamma_{gr}$ controller and any shape controller would compete for actuator authority through the same PF circuits. Pesamosca et~al.~\cite{pesamosca2022_tcv} addressed the analogous problem on TCV by using two different optimal coil combinations: a fast combination for vertical stabilization and a slow combination compatible with shape control. The present single-QP architecture does not make this timescale separation explicitly. Recent work on TCV has also implemented a proximity controller based on real-time $\gamma_{gr}$ estimation,  finding that the radial actuation channel maintained the growth rate effectively at a set reference value,  while a shape-based actuation channel showed more limited performance consistent with the many-to-one nature of the $\gamma_{gr}$-shape relationship discussed above.

For deployment in an integrated control system,  the $\gamma_{gr}$ controller presented here would operate as the fast inner loop of a cascade architecture. A slower outer loop updating gap references,  $\kappa$ targets,  or direct shape parameters at ${\sim}10$ to $60$~Hz would provide the shape regulation that the present QP omits. The interface requirement is an actuator authority budget: the outer loop must leave sufficient PF voltage headroom for the inner $\gamma_{gr}$ controller to respond to transients. Quantifying this budget for the ARC~V3A coil geometry is a subject for future work.

\subsection{Controllability boundaries}

The simulation matrix identifies three distinct failure mechanisms.

The first is voltage-authority failure (A10,  B2). When the initial growth rate exceeds ${\sim}10$~s$^{-1}$,  the 500~V limit prevents the out-vessel coils from driving current fast enough. The critical growth rate for a given voltage limit can be estimated as $\gamma_{z, \mathrm{crit}} \sim V_{\max}/(L_{\mathrm{PF}} \cdot \Delta Z_{\mathrm{margin}})$,  though the actual boundary depends on coil geometry and passive-structure coupling. This is the same mechanism Nishio et~al.~\cite{nishio1993} identified for ITER CDA at $\kappa=2.0$,  albeit at substantially higher growth rates (${\sim}10$ to $10^3$~s$^{-1}$ in their analysis) due to the higher elongation and deteriorated plasma conditions considered. The moderate ARC~V3A growth rates ($1.8$ to $11.2$~s$^{-1}$ across the operating space) reflect the lower target elongation ($\kappa_{\mathrm{sep}}=1.80$) relative to Nishio's $\kappa=2.0$ case. At growth rates approaching the controllability boundary,  the failure is compounded by accumulated sensitivity staleness (Section~\ref{sec:controller}): over the ${\sim}89$ control cycles within one growth time at $\gamma_{gr} = 11.2$~s$^{-1}$,  the evolving equilibrium causes $\bm{S}_\gamma$ to drift from the initial linearization,  so the controller is simultaneously voltage-limited and progressively misdirected.

The second is compound degradation failure (B7,  B8). Each individual degradation (reduced voltage,  increased latency,  slower power supply) may be tolerable,  but combinations erode the stability margin from multiple directions simultaneously. This has direct implications for system design: specifying actuator and diagnostic requirements independently,  without considering their interaction,  can produce designs that are individually adequate but collectively insufficient.

The third is surrogate extrapolation failure (C4,  C5,  C6). Large disturbances push the equilibrium outside the training distribution,  causing the surrogate prediction error and sensitivity accuracy to degrade simultaneously. The controller may retain approximately correct gradient direction but lose magnitude accuracy,  producing oscillation or sluggish response. This is an ML-specific limitation that distinguishes the present approach from physics-based controllers: a classical PD controller degrades gracefully under model uncertainty,  while a surrogate-based controller could fail abruptly at the training boundary.

\section{Conclusions}\label{sec:conclusions}

We have presented a model-based feedback controller that regulates the vertical instability growth rate $\gamma_{gr}$ as the primary controlled variable in double-null ARC~V3A plasmas,  using only out-vessel PF coils. The approach combines a dual-branch ML surrogate trained on non-rigid linearized FGE stability calculations with $P'$ and $TT'$ profile inputs,  automatic differentiation for state dependent actuator sensitivities,  and a constrained QP for coil-voltage allocation across multiple control objectives.

Across 24 closed-loop simulation cases spanning equilibrium variations,  actuator degradations,  and transient disturbances,  the controller meets the target criteria in 12 cases,  achieves marginal performance in 8,  and fails in 4. The failures identify the boundaries of out-vessel controllability: initial growth rates above ${\sim}10$~s$^{-1}$ (from high $\kappa$ combined with peaked $\ell_i$) exceed the 500~V voltage authority,  combined actuator-diagnostic degradations erode stability margins from multiple directions,  and large disturbances that push the equilibrium beyond the surrogate training distribution cause simultaneous prediction and sensitivity degradation. These findings are consistent with the early out-vessel controllability analysis of Nishio et~al.~\cite{nishio1993} and the SPARC vertical stability assessment of Nelson et~al.~\cite{nelson2024_sparc_vs}.

The controller does not include explicit boundary shape regulation. Separatrix elongation remained within $\pm 0.02$ of the target in the baseline case,  but this is a consequence of the weak coupling between PF5 differential-mode commands and the shape setting field harmonics not a guarantee of the control architecture. Sustained operation or scenario evolution would require a separate shape control layer,  and the actuator authority budget between $\gamma_{gr}$ and shape regulation is an open design question for the ARC~V3A coil geometry.

The principal open questions are robustness to diagnostic noise and reconstruction errors,  systematic characterization of surrogate gradient accuracy,  generalization of the surrogate beyond the training distribution,  real-time implementation on PCS hardware with deterministic timing guarantees,  and design of a fallback hierarchy for safe degradation. Hardware-in-the-loop testing and deployment on a present-day tokamak are the necessary next steps. We expect the structured model-based  framework physics-informed surrogate,  gradient-based sensitivities,  constrained optimization to transfer more reliably across devices than end-to-end learned controllers,  though this hypothesis remains to be tested experimentally.

\section*{Acknowledgments}
This work is supported by Commonwealth Fusion Systems (MHD-RPP023). Simulations were performed on MIT-PSFC computing resources. We acknowledge the MEQ development team at EPFL for their contributions to the underlying physics modeling framework.

\bibliographystyle{elsarticle-num}
\bibliography{references}

\end{document}